\documentclass{article}

\usepackage{amsmath, amsthm, amssymb}
\usepackage{graphicx,multirow}
\usepackage{verbatim}
\usepackage{natbib}
\usepackage{caption}
\usepackage{subcaption}
\usepackage{fancyvrb}
\usepackage{enumerate}
\usepackage{relsize}
\usepackage{float}
\usepackage{setspace}

\usepackage{hyperref}
\usepackage[margin=1.5in]{geometry}
\hypersetup{colorlinks,citecolor=blue,urlcolor=blue,linkcolor=blue}

\usepackage{stefan_tex}
\graphicspath{{./figures/}}

\newcommand{\acal}{\mathcal{A}}

\theoremstyle{plain}
\newtheorem{prop}{Proposition}

\newtheorem{theo}[prop]{Theorem}

\theoremstyle{definition}
\newfloat{algbox}{tbp}{lop}
\newtheorem{alg}{Procedure}
\newcommand{\myalg}[3]{
\begin{center}
\fbox{
\parbox{0.95\textwidth}{
\begin{alg}\label{#1}{\textsc{ #2}}
\vspace{.1cm}\\ #3 
\end{alg}
}}
\end{center}
}

\theoremstyle{remark}

\author{Guido Imbens \\ \texttt{imbens@stanford.edu}
\and
Stefan Wager \\ \texttt{swager@stanford.edu}}

\date{Current version \ifcase\month\or
January\or February\or March\or April\or May\or June\or
July\or August\or September\or October\or November\or December\fi \ \number%
\year\ \  }

\title{Optimized Regression Discontinuity Designs\footnote{We are grateful for helpful
comments from Timothy Armstrong, Max Farrell, Michal Koles\'ar, Christoph Rothe, Cun-Hui Zhang,
seminar participants at Berkeley, Stanford and the University of Chicago, as well as the editor and four anonymous referees.}}

\begin{document}

\maketitle

\begin{abstract}
The increasing popularity of regression discontinuity methods for causal inference in observational
studies has led to a proliferation of different estimating strategies, most of which involve first fitting
non-parametric regression models on both sides of a treatment assignment boundary and then reporting plug-in estimates
for the effect of interest. In applications, however, it is often difficult to tune the non-parametric
regressions in a way that is well calibrated for the specific target of inference; for example, the model with
the best global in-sample fit may provide poor estimates of the discontinuity parameter.
In this paper, we propose an alternative method for estimation and  statistical inference in regression discontinuity designs
that uses numerical convex optimization to directly obtain the finite-sample-minimax linear estimator for the
regression discontinuity parameter, subject to bounds on the second derivative of the conditional response function.
Given a bound on the second derivative, our proposed method is fully data-driven,
and provides uniform confidence intervals for the regression discontinuity parameter
with both discrete and continuous running variables. The method also naturally extends to the
case of multiple running variables.
\end{abstract}



\section{Introduction}

Regression discontinuity designs, first developed in the sixties \citep*{thistlethwaite1960regression},
often allow for simple and transparent identification of treatment effects
from observational data \citep*{hahn2001identification,imbens2008regression,trochim1984research},
and their statistical properties have been the subject of recent interest
\citep*{armstrong2016optimal,calonico2014robust,cheng1997automatic,kolesar2016inference}.
The sharp regression discontinuity design assumes a treatment assignment generated by a running variable $X \in \RR^k$,
such that individuals get treated if and only $X \in \acal$ for some set $\acal \subset \RR^k$.
For example, in epidemiology, $X \in \RR$ could be a severity index and patients are
assigned a medical intervention whenever $X \geq c$ for some threshold $c$ (i.e., $\acal = \cb{x \in \RR : x \geq c}$);
or, in political science, $X \in \RR^2$ could denote the latitude and longitude of a household,
and $\acal$ could be an administrative region that has enacted a specific policy.

Given appropriate assumptions, we can identify a causal effect by comparing subjects $i$ with $X_i$
barely falling within the treatment region $\acal$ to those with $X_i$ just outside of it.
Variants of this identification strategy have proven to
be useful in education \citep{angrist1999using,black1999better,jacob2004remedial},
political science \citep{caughey2011elections, keele2014geographic, lee2008randomized},
criminal justice \citep{berk1983capitalizing},
program evaluation \citep{lalive2008extended,ludwig2007does},
and other areas.
As discussed in more detail below, standard methods for inference in the regression discontinuity design
rely on plugin-in estimates from local linear regression.

In this paper, motivated by a large literature on minimax linear estimation
\citep{armstrong2016optimal,cai2003note,donoho1994statistical,donoho1991geometrizing,
ibragimov1985nonparametric,johnstone2011gaussian,juditsky2009nonparametric},
we study an alternative approach based on directly minimizing finite sample error bounds via numerical optimization,
under an assumption that the second derivative of the response surface is bounded away from the
boundary of the treatment region.\footnote{Of these papers, our work is most closely
related to that of \citet{armstrong2016optimal}, who explicitly consider minimax linear estimation in the
regression discontinuity model for an ``approximately linear'' model in the sense of \citet{sacks1978linear}
that places restrictions on second differences relative to the response surface at the threshold.
In contrast, we assume bounded second derivatives away from the threshold.
An advantage of their approach is that it allows for a closed form solution. However, a disadvantage is that
they allow for discontinuities in the response surface away from the threshold, which implies that
given the same value for our bound on the second derivative and their bound on
second differences, our confidence intervals can be substantially shorter (moreover, allowing for
discontinuities in the response surface does not seem conceptually attractive given that the assumption of
continuity of the conditional expectation at the threshold is fundamental to the regression discontinuity design).
We discuss this comparison further in Section \ref{sec:literature}.}
This approach has several advantages relative to local regression. Our estimator is well defined regardless of the
shape of the treatment region $\acal$, whether it be a half line as in the standard univariate regression discontinuity
specification or an oddly shaped region as might appear with a geographic regression discontinuity; moreover,
our inference is asymptotically valid for both discrete and continuous running variables. Finally, even with
univariate designs, our approach strictly dominates local linear regression in terms of minimax mean-squared error.

For simplicity of exposition, we start by presenting our method in the context of classical univariate regression
discontinuity designs with a single treatment cutoff, i.e., with $X_i \in \RR$ and $\acal = \cb{x \in \RR: x \geq c}$;
a solution to the more general problem will then follow by direct extension.
A software implementation, \texttt{optrdd} for \texttt{R}, is available on \texttt{CRAN} and from
\url{github.com/swager/optrdd}.

\subsection{Optimized Inference with Univariate Discontinuities}
\label{sec:intro_uni}

We have access to $i=1, \, ..., \, n$ independent pairs $(X_i, \, Y_i)$ where $X_i \in \RR$
is the running variable and $Y_i \in \RR$ is our outcome of interest; the treatment is assigned as $W_i = \ind\p{\cb{X_i \geq c}}$.
Following the potential outcomes model \citep{imbens2015causal,neyman1923applications,rubin1974estimating},
we posit potential outcomes $Y_i(w)$, for $w \in \cb{0, \, 1}$ corresponding to the outcome subject $i$ would have
experienced had they received treatment $w$, and define the conditional average treatment effect $\tau(x)$
in terms of the conditional response functions $\mu_w(x)$:
\begin{equation}
\label{eq:cate}
\mu_w(x) = \EE{Y_i(w) \cond X_i = x}, \ \ \tau(x) = \mu_1(x) - \mu_0(x).
\end{equation}
Provided the functions $\mu_w(x)$ are both continuous at $c$,
the regression discontinuity identifies the conditional average treatment effect at the
threshold $c$,\footnote{In the fuzzy regression discontinuity design where
the probability of receiving the treatment changes discontinuously at $x=c$, but not necessarily from
zero to one, the estimand can be written as the ratio of two such differences.
The issues we address in this paper also arise in that setting,
and the present discussion extends naturally to it;
see Section \ref{sec:discussion} for a discussion.}
\begin{equation}
\label{eq:identification}
\tau(c) = \lim_{x \downarrow c} \EE{Y_i \cond X_i = x} - \lim_{x \uparrow c} \EE{Y_i \cond X_i = x}.
\end{equation}
Given this setup, local linear regression is a popular strategy for estimating $\tau(c)$
\citep{hahn2001identification, porter2003estimation}:
\begin{equation}
\label{eq:llr}
\htau = \argmin\cb{\frac{1}{n h_n} \sum_{i = 1}^n K\p{\abs{\Delta_i}\big/{h_n}} \p{Y_i - a - \tau W_i - \beta_- \p{\Delta_i}_- - \beta_+ \p{\Delta_i}_+}^2},
\end{equation}
where $K(\cdot)$ is some weighting function, $h_n$ is a bandwidth,
$\Delta_i = X_i - c$, and $a$ and $\beta_{\pm}$ are nuisance parameters.
When we do not observe data right at the boundary $c$ (e.g., when $X_i$ has
discrete support), then $\tau(c)$ is not point identified; however, given smoothness
assumptions on $\mu_w(x)$, \citet{kolesar2016inference} propose an approach to local linear
regression that can still be used to construct partial identification intervals for $\tau(c)$
in the sense of \citet{imbens2004confidence} (see Section \ref{sec:inference} for a discussion).

The behavior of regression discontinuity estimation via local linear regression is fairly
well understood. When the running variable $X$ is continuous (i.e., $X$ has a continuous positive density at $c$)
and $\mu_w(x)$ is twice differentiable with a bounded second derivative in a neighborhood of $c$,
\citet*{cheng1997automatic} show that the triangular kernel $K(t) = (1 - t)_+$ minimizes worst-case
asymptotic mean-squared error among all possible choices of $K$,
\citet{imbens2011optimal} provide a data-adaptive choice of $h_n$ to minimize the mean-squared
error of the resulting estimator, and
\citet*{calonico2014robust} propose a method for removing bias effects due to the curvature of $\mu_w(x)$
to allow for asymptotically unbiased estimation.
Meanwhile, given a second-derivative bound $\abs{\mu_{w}''(x)} \leq B$,
\citet{armstrong2016optimal} and \citet{kolesar2016inference} construct
confidence intervals centered at the local linear estimator \smash{$\htau$} that attain uniform asymptotic coverage,
even when the running variable $X$ may be discrete.

Despite its ubiquity, however, local linear regression still has some shortfalls. First of all, under the  bounded
second derivative assumption often used to justify local linear regression (i.e., that $\mu_{w}(x)$ is twice
differentiable and $\abs{\mu_{w}''(x)} \leq B$ in a neighborhood of $c$), local linear regression is not the minimax
optimal linear estimator for $\tau(c)$---even with a continuous running variable.
Second, and perhaps even more importantly, all the motivating theory for local linear regression relies on $X$
having a continuous distribution; however, in practice, $X$ often has a discrete distribution with a modest number of points of support.
 When the running variable is discrete there is no compelling reason to expect local linear regression to be particularly
effective in estimating the causal effect of interest.\footnote{One practical inconvenience that can arise in local
linear regression with discrete running variables is that, if we use a data-driven rule to pick the bandwidth
$\hat{h}$ (e.g., the one of \citet{imbens2011optimal}), we may end up with no data inside the specified
range (i.e., there may be no observations with $|X_i - c| \leq h$); the practitioner is then forced to select a different
bandwidth ad-hoc. Ideally, methods for regression discontinuity analysis should be fully data-driven, even
when $X$ is discrete.}
In spite of these limitations, local linear regression is still the method of choice, largely because of its intuitive appeal.

As discussed above, the goal of this paper is to show that we can systematically do better.
Regardless of the shape of the kernel $K(\cdot)$ in \eqref{eq:llr}, local linear regression yields a
\emph{linear estimator}\footnote{We note the unfortunate terminological overlap between ``local linear
regression'' estimators of $\tau$ and ``linear'' estimators of type \smash{$\htau = \sum_{i = 1}^n \hgamma_i Y_i$}.
The word linear in these two contexts refers to different things.
All ``local linear regression'' estimators are ``linear'' in the latter sense, but not vice-versa.}
for $\tau$, i.e., one of the form \smash{$\htau = \sum_{i = 1}^n \hgamma_i Y_i$}
for weights \smash{$\hgamma_i$} that depend only on the distances $X_i - c$. Here, we
find that if we are willing to rely on numerical optimization tools, then \emph{minimax linear} estimation of $\tau(c)$,
i.e., minimax among all estimators of the form \smash{$\htau = \sum_{i = 1}^n \hgamma_i Y_i$} conditionally on $X_i$,
is both simple and methodologically transparent under natural assumptions on the regularity of $\mu_{(w)}(x)$.
If we know that \smash{$\Var{Y_i \cond X_i} = \sigma_i^2$} and that \smash{$\abs{\mu''_w(x)} \leq B$},
we propose estimating $\tau(c)$ as follows:
\begin{equation}
\label{eq:estimator}
\begin{split}
&\htau = \sum_{i = 1}^n \hgamma_i Y_i, \ \
\hgamma = \argmin_{\gamma} \cb{\sum_{i = 1}^n \gamma_i^2 \sigma_i^2 
 + I_B^2(\gamma)}, \\
& I_B\p{\gamma} := \sup_{\mu_0(\cdot),\mu_1(\cdot)}\cb{\sum_{i = 1}^n \gamma_i \mu_{W_i}(X_i) - \p{\mu_1(c) - \mu_0(c)} : \abs{\mu''_w(x)} \leq B \text{ for all } w, \, x}.
\end{split}
\end{equation}
Because this estimator is minimax among the class of linear estimators, it is at least as accurate as any
local linear regression estimator in a minimax sense over all problems with $\Var{Y_i \cond X_i} = \sigma_i^2$ and
$\abs{\mu''_w(x)} \leq B$. For further discussion of related estimators, see \citet{cai2003note},
\citet{donoho1994statistical}, \citet{donoho1991geometrizing}, and
\citet{juditsky2009nonparametric}.

When $X_i$ has a discrete distribution, the parameter $\tau(c)$ is usually not point identified because
there may not be any observations $X_i$ in a small neighborhood of $c$.
However, we can get meaningful partial identification of $\tau(c)$ thanks to our bounds on the second derivative
of $\mu_w(x)$. Moreover, because our approach controls for bias in finite samples, the estimator
\eqref{eq:estimator} is still justified in the partially identified setting and, as discussed further in
Section \ref{sec:inference}, provides valid confidence intervals for $\tau(c)$ in
the sense of \citet{imbens2004confidence}.
We view the fact that our estimator can seamlessly move between the point and partially identified
settings as an important feature.

To motivate the above estimator qualitatively, note that the first term in the minimization problem corresponds to the conditional
variance of \smash{$\htau$} given \smash{$\cb{X_i}$}, while the second term is the worst-case conditional
squared bias given that \smash{$|\mu''_w(x)| \leq B$}; thus, given our regularity assumptions, the estimator
\smash{$\htau$} minimizes the worst-case conditional mean-squared error among all linear estimators.
Because no constraints are placed on $\mu_w(c)$ or $\mu'_w(c)$, the optimization
in \eqref{eq:estimator} also automatically enforces the constraints
\smash{$\sum_i W_i\hgamma_i=1$},
\smash{$\sum_i (1 - W_i)\hgamma_i=-1$},
\smash{$\sum_i W_i (X_i - c)\hgamma_i=0$}, and
\smash{$\sum_i (1 - W_i)(X_i - c)\hgamma_i=0$}.\footnote{At values of $\gamma$ for which 
these constraints are not all satisfied, we can choose $\mu_1(x)$ and $\mu_0(x)$ with second derivative
bounded by $B$ such as to make the conditional bias arbitrarily bad, i.e., $I_B(\gamma) = +\infty$.
Thus, the solution \smash{$\hgamma$} to \eqref{eq:estimator} must satisfy the constraints.}
This is a convex program, and can be efficiently solved using readily available software described
in, e.g., \citet{boyd2004convex}.

Although this estimator depends explicitly on knowledge of $\sigma_i^2$ and $B$, we note that all practical
methods for estimation in the regression discontinuity model, including 
\citet{calonico2014robust}, \citet{imbens2011optimal} and \citet{kolesar2016inference},
require estimating related quantities in order to tune the algorithm. Then, once estimators for these parameters have been
specified, the procedure \eqref{eq:estimator} is fully automatic---in particular, there is no need to ask whether
the running variable is discrete or continuous, as the optimization is conditional on $\cb{X_i}$---whereas the
baseline procedures still have other choices to make, e.g., what weight function $K(\cdot)$ to use, or whether
to debias the resulting $\smash{\htau}$-estimator.
See Sections \ref{sec:tuning} and \ref{sec:multi} for discussions on how to choose $B$ in practice.

\setlength{\tabcolsep}{1pt}
\begin{figure}[t]
\begin{center}
\begin{tabular}{cccc}
\includegraphics[width=0.24\textwidth, trim=7mm 7mm 7mm 7mm]{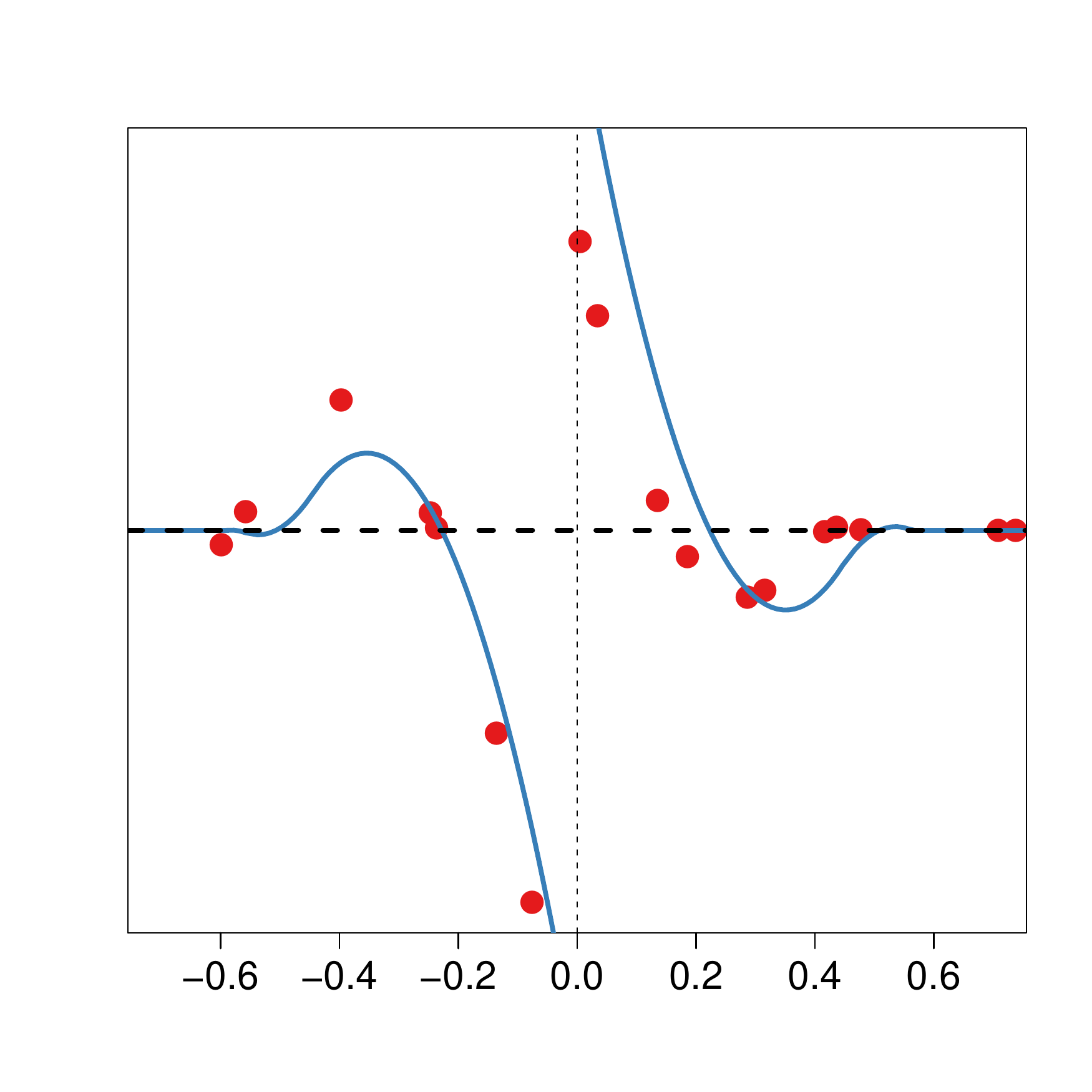} &
\includegraphics[width=0.24\textwidth, trim=7mm 7mm 7mm 7mm]{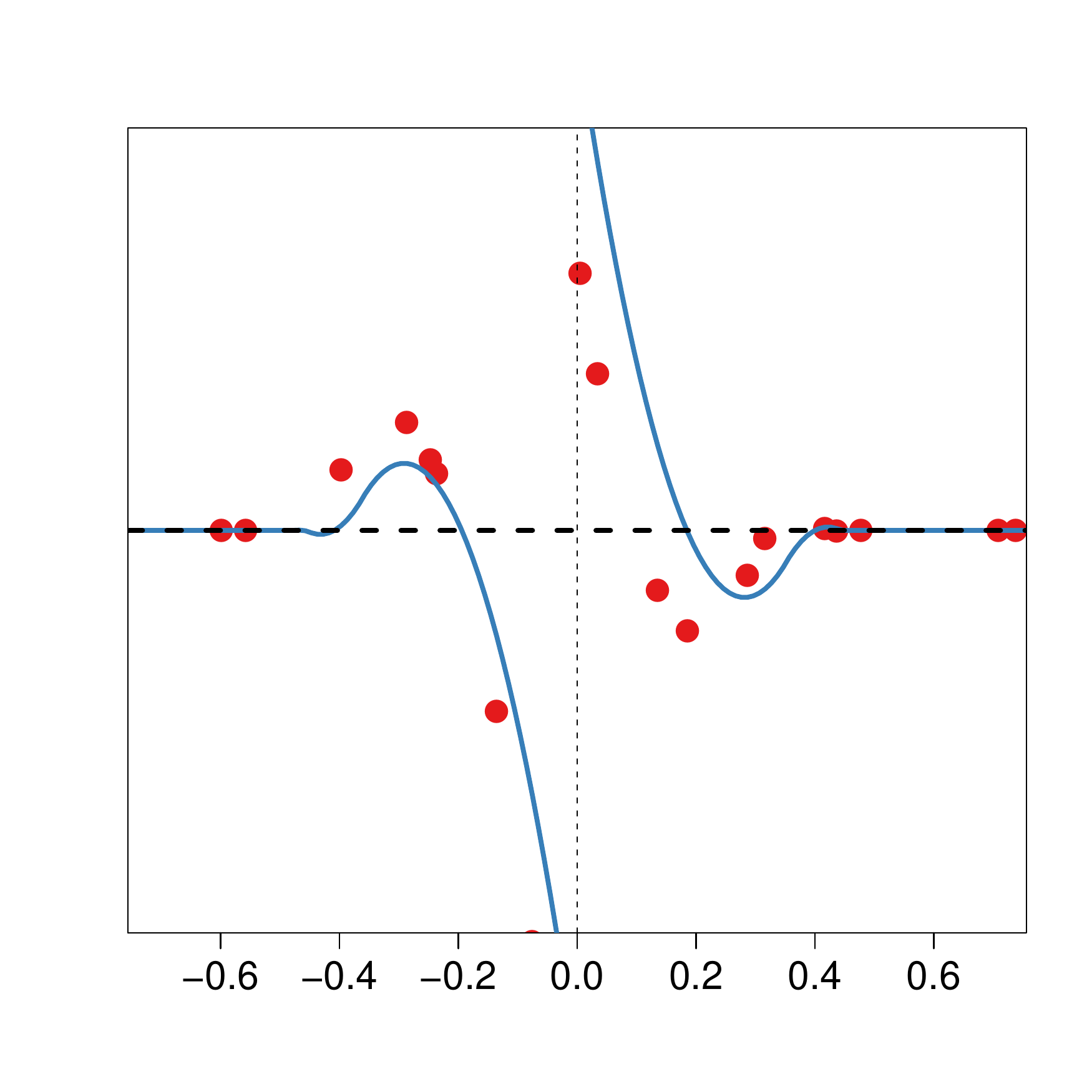} &
\includegraphics[width=0.24\textwidth, trim=7mm 7mm 7mm 7mm]{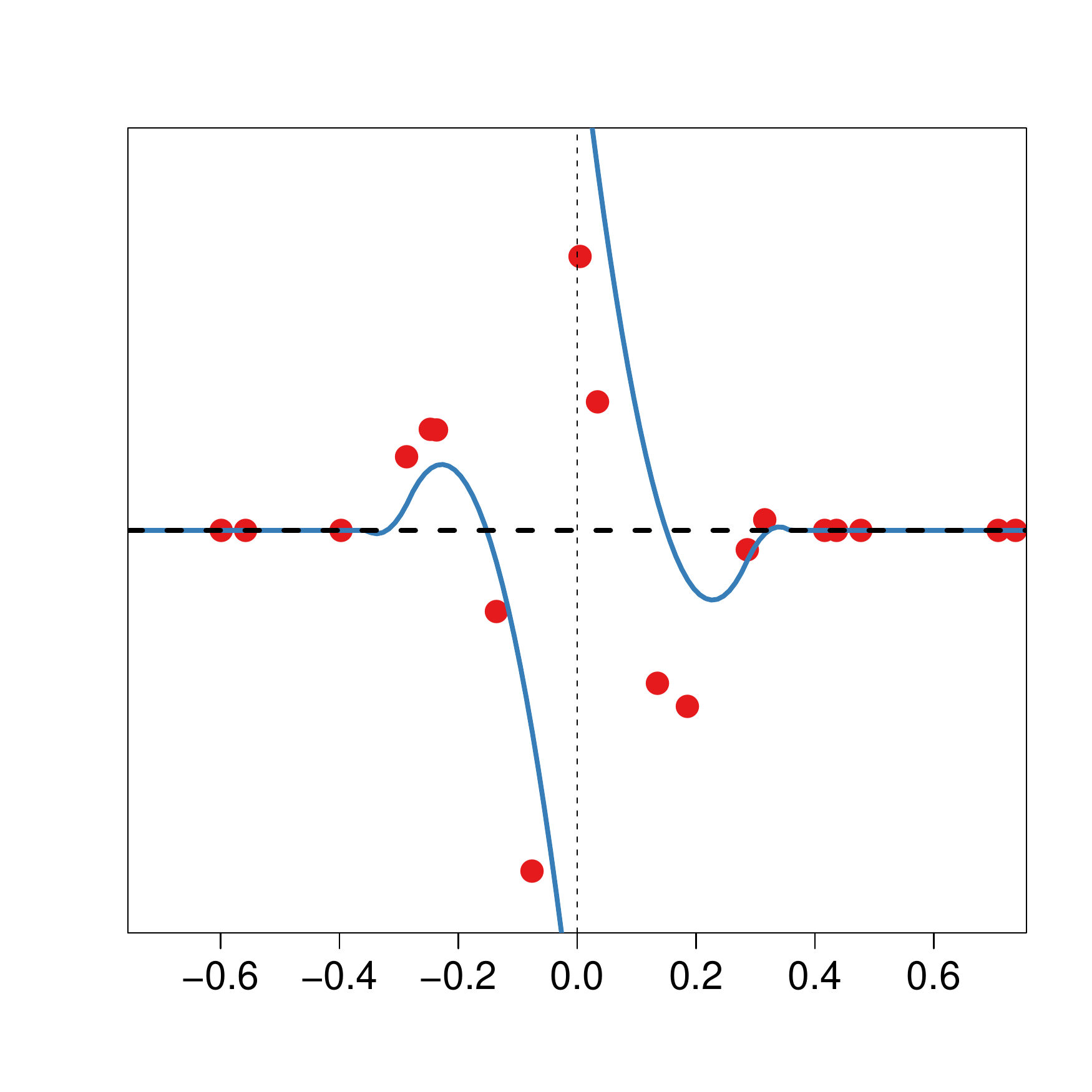} &
\includegraphics[width=0.24\textwidth, trim=7mm 7mm 7mm 7mm]{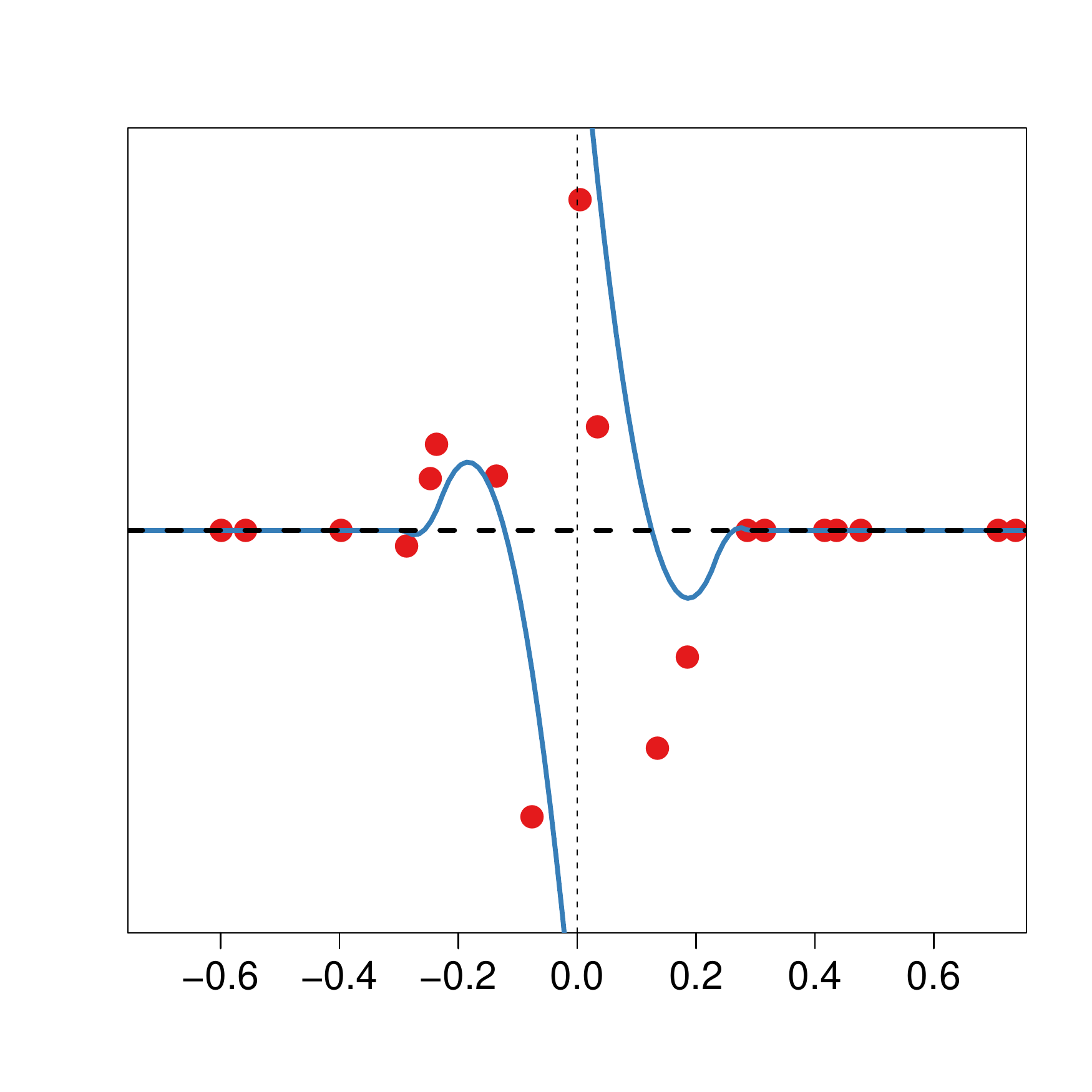} \\
$n = 1,000$ & $n = 3,000$ & $n = 9,000$ & $n = 27,000$
\end{tabular}
\caption{Optimized regression discontinuity design obtained via \eqref{eq:estimator}, for
different values of $n$ and two different $X$ distributions. The red dots show the learned
weighting function in a case where the running variable $X$ is discrete, and different support points
are sampled with different probabilities (the probability mass function is shown in the left panel of Figure \ref{fig:mse_cmp}); the
blue line shows $\gamma(X_i)$ for standard Gaussian $X$. We plot $n^{4/5} \hgamma_i$, motivated by the
fact that, with a continuous running variable, the optimal bandwidth for local linear regression
scales as $h_n \sim n^{-1/5}$. The weights $\hgamma_i$ were computed with $B = 5$ and $\sigma^2 = 1$.}
\label{fig:intro}
\end{center}
\end{figure}
\setlength{\tabcolsep}{6pt}

\begin{figure}[t]
\centering
\begin{tabular}{cc}
\includegraphics[width=0.45\textwidth, trim=7mm 7mm 7mm 7mm]{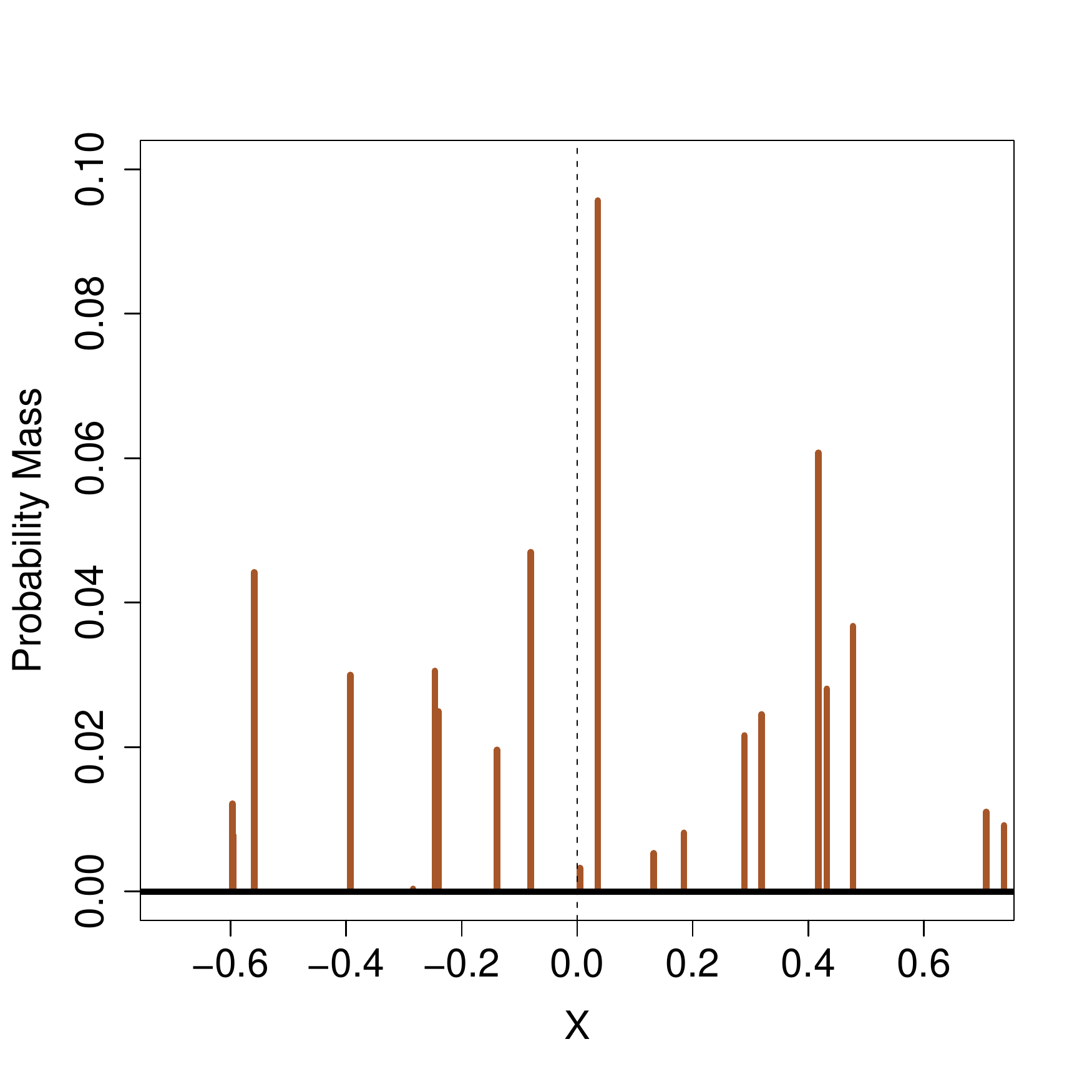} &
\includegraphics[width=0.45\textwidth, trim=7mm 7mm 7mm 7mm]{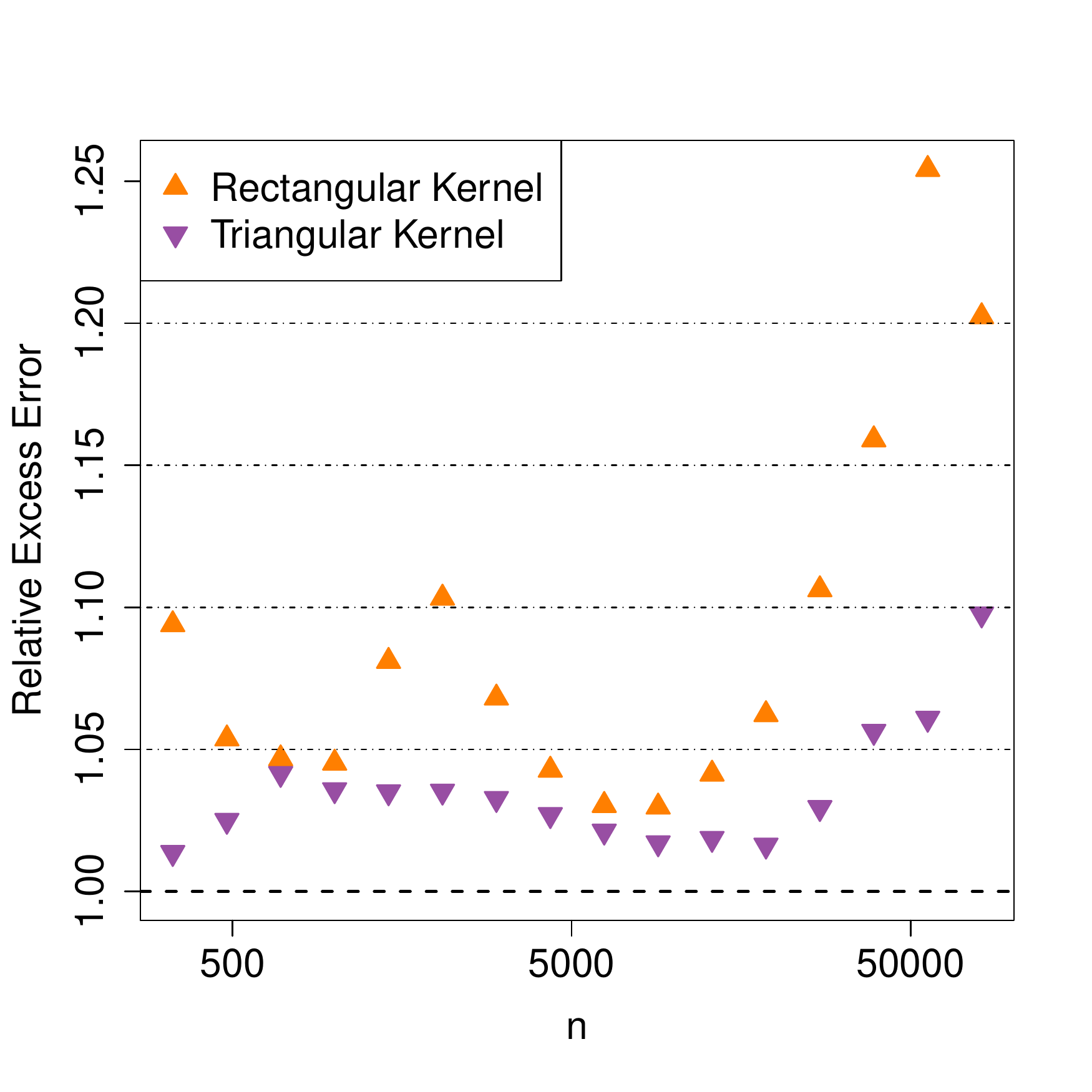}
\end{tabular}
\caption{Left panel: Probability mass function underlying the example in Figure \ref{fig:intro}
and the right panel of the present figure.
Right panel: Comparison of our procedure \eqref{eq:estimator} with local linear regression,
both using a rectangular ($K(t) = 1\p{\cb{t \leq 1}}$) and triangular ($K(t) = (1 - t)_+$) kernel.
We compare methods in terms of their worst-case mean-squared error conditional on $\cb{X_i}$;
for local linear regression, we always chose the bandwidth to make this quantity as small as possible.
We depict performance relative to our estimator \eqref{eq:estimator}.}
\label{fig:mse_cmp}
\end{figure}

Figure \ref{fig:intro} compares the weights  $\hgamma_i$ obtained via \eqref{eq:estimator} in two
different settings: one with a discrete, asymmetric running variable $X$ depicted in the left panel of
Figure \ref{fig:mse_cmp}, and the other with a standard Gaussian running variable.
We see that, for $n = 1,000$, the resulting weighting functions look fairly similar, and are
also comparable to the implicit weighting function generated by local linear regression with a triangular kernel.
However, as $n$ grows and the discreteness becomes more severe, our method changes both the shape and the scale of the
weights, and the discrepancy between the optimized weighting schemes for discrete
versus continuous running variables becomes more pronounced.

In the right panel of Figure \ref{fig:mse_cmp}, we also compare the worst-case conditional mean-squared error of our method
relative to that of optimally tuned local linear regression, both with a rectangular and triangular kernel.
For the smallest sample size we consider, $n = 333$, the discreteness of the running variable has a fairly
mild effect on estimation and---as one might have expected---the triangular kernel is noticeably better than
the rectangular kernel, while our method is slightly better than the triangular kernel. However, as the sample
size increases, the performance of local linear regression relative to our method ebbs and flows rather
unpredictably.\footnote{As a matter of intellectual curiosity, it is intriguing to ask whether
there exist discrete distributions for which the rectangular kernel may work substantially better than the triangular
kernel, or whether additional algorithmic tweaks---such as using different bandwidths on different sides of the
threshold---may have helped (in the above example, we used the same bandwidth for local linear regression
on both sides of the boundary). However, from a practical perspective, the estimator \eqref{eq:estimator} removes
the need to consider such questions in applied data analysis, and automatically adapts to the structure
of the data at hand.}

\subsection{Optimized Inference with Generic Discontinuities}
\label{sec:intro_multi}

The methods presented above extend naturally to the general case, where $X_i \in \RR^k$ may be
multivariate and $\acal$ is unrestricted. 
The problem of regression discontinuity inference with multiple running variables is
considerably richer than the corresponding problem with a single running variable,
because an investigator could now plausibly hope to identify many different treatment
effects along the boundary of the treated region $\acal$.
Most of the existing literature on this setup, including \citet{papay2011extending},
\citet{reardon2012regression} and \citet{wong2013analyzing}, have focused on these
questions of identification, while using some form of local linear regression for
estimation.

In the multivariate case, however, questions about how to tune local linear regression
are exacerbated, as the problems of choosing the kernel function $K(\cdot)$ and the
bandwidth $h$ are now multivariate. Perhaps for this reason, it is still popular to
use univariate methods to estimate treatment effects in the multivariate
setting by, e.g., using shortest distance to the boundary of the treatment region $\acal$ as a univariate
running variable \citep{black1999better}, or only considering a subset of the
data where univariate methods are appropriate \citep{jacob2004remedial,matsudaira2008mandatory}.

Here, we show how our optimization-based method can be used to side-step the problem of choosing
a multivariate kernel function by hand. In addition to providing a simple-to-apply
algorithm, our method lets us explicitly account for the curvature of the
mean-response function $\mu_w(x)$ for statistical inference, thus strengthening
formal guarantees relative to prior work. 

Relative to the univariate case, the multivariate case has two additional subtleties we need to address. First,
in \eqref{eq:estimator} it is natural to impose a constraint $|\mu_w''(x)| \leq B$ to ensure smoothness; in
the multivariate case, however, we have more choices to make. For example, do we constrain $\mu_w(x)$ to
be an additive function, or do we allow for interactions? Here, we opt for the more flexible specification, and
simply require that $\Norm{\nabla^2 \mu_w(x)} \leq B$, where $\Norm{\cdot}$ denotes the operator norm
(i.e., the largest absolute eigenvalue of the second derivative).

Moreover, as emphasized by \citet{papay2011extending}, whereas the univariate design only enables us to identify
the conditional average treatment effect at the threshold $c$, the multivariate design enables us to potentially
identify a larger family of treatment effect functionals.
Here, we focus on the following two causal estimands. First, writing $c$ for a focal point of interest,
let $\htau_c = \sum_{i = 1}^n \hgamma_{c,i} Y_i$ with
\begin{equation}
\label{eq:estimator_2d_pt}
\hgamma_c = \argmin_{\gamma} \cb{\sum_{i = 1}^n \gamma_i^2 \sigma_i^2 + \p{\sup_{\Norm{\nabla^2\mu_w(x)} \leq B} \cb{\sum_{i = 1}^n \gamma_i \mu_{W_i}(X_i) - \p{\mu_1(c) - \mu_0(c)}}}^2}
\end{equation}
denote an estimator for the conditional average treatment effect at $c$. The upside of this approach is that
it gives us an estimand that is easy to interpret; the downside is that, when curvature is non-negligible,
\eqref{eq:estimator_2d_pt} can effectively only make use of data near the specified focal point $c$, thus resulting
in relatively long confidence intervals.

In order to potentially improve precision, we also study weighted conditional average treatment effect estimation with
weights greedily chosen such as to make the inference as precise as possible:
In the spirit of \citet{crump2009dealing}, \citet{li2016balancing} or \citet{robins2008higher}, we consider
$\htau_* = \sum_{i = 1}^n \hgamma_{*,i} Y_i$, with
\begin{equation}
\label{eq:estimator_2d_greedy}
\begin{split}
\hgamma_* &= \argmin_{\gamma} \cb{\sum_{i = 1}^n \gamma_i^2 \sigma_i^2 + \p{\sup_{\Norm{\nabla^2\mu_0(x)} \leq B} \cb{\sum_{i = 1}^n \gamma_i \mu_{0}(X_i)}}^2 : \sum_{i = 1}^n \gamma_i W_i = 1}.
\end{split}
\end{equation}
In other words, we seek to pick weights $\gamma_i$ that are nearly immune to bias
due to curvature of the baseline response surface $\mu_{0}(x)$.
By construction, this estimator satisfies
\begin{equation}
\label{eq:weighted_estimand}
\abs{\EE{\htau_* \cond \cb{X_i}} - \btau(\hgamma_*)} \leq \sup_{\Norm{\nabla^2\mu_0(x)} \leq B} \cb{\sum_{i = 1}^n \hgamma_{*,i} \mu_{0}(X_i)}, \ \ \ \btau(\hgamma_*) :=  \sum_{i = 1}^n W_i \, \hgamma_{*,i} \tau(X_i).
\end{equation}
Because  $\sum W_i \hgamma_{*,i} = 1$, we see that $\btau(\hgamma_{*})$ is in fact a weighted average
of the conditional average treatment effect function $\tau(\cdot)$ over the treated sample. If we ignored the curvature of
$\tau(\cdot)$, we could interpret $\htau_*$ as an estimate for the conditional average treatment effect
at $x_* = \sum \hgamma_{*,i} W_i X_i$.

In some cases, it is helpful to consider other interpretations of the estimand underlying \eqref{eq:estimator_2d_greedy}.
It we are willing to assume a constant treatment effect $\tau(x) = \tau$, then \smash{$\btau = \tau$}, and \smash{$\htau_*$}
is the minimax linear estimator for $\tau$. Relatedly, we can always use the confidence intervals from Section \ref{sec:inference}
built around \smash{$\htau_*$} to test the global null hypothesis $\tau(x) = 0$ for all $x$.

\begin{figure}[t]
\centering
\begin{tabular}{cc}
\includegraphics[width=0.45\textwidth, trim=7mm 7mm 7mm 7mm]{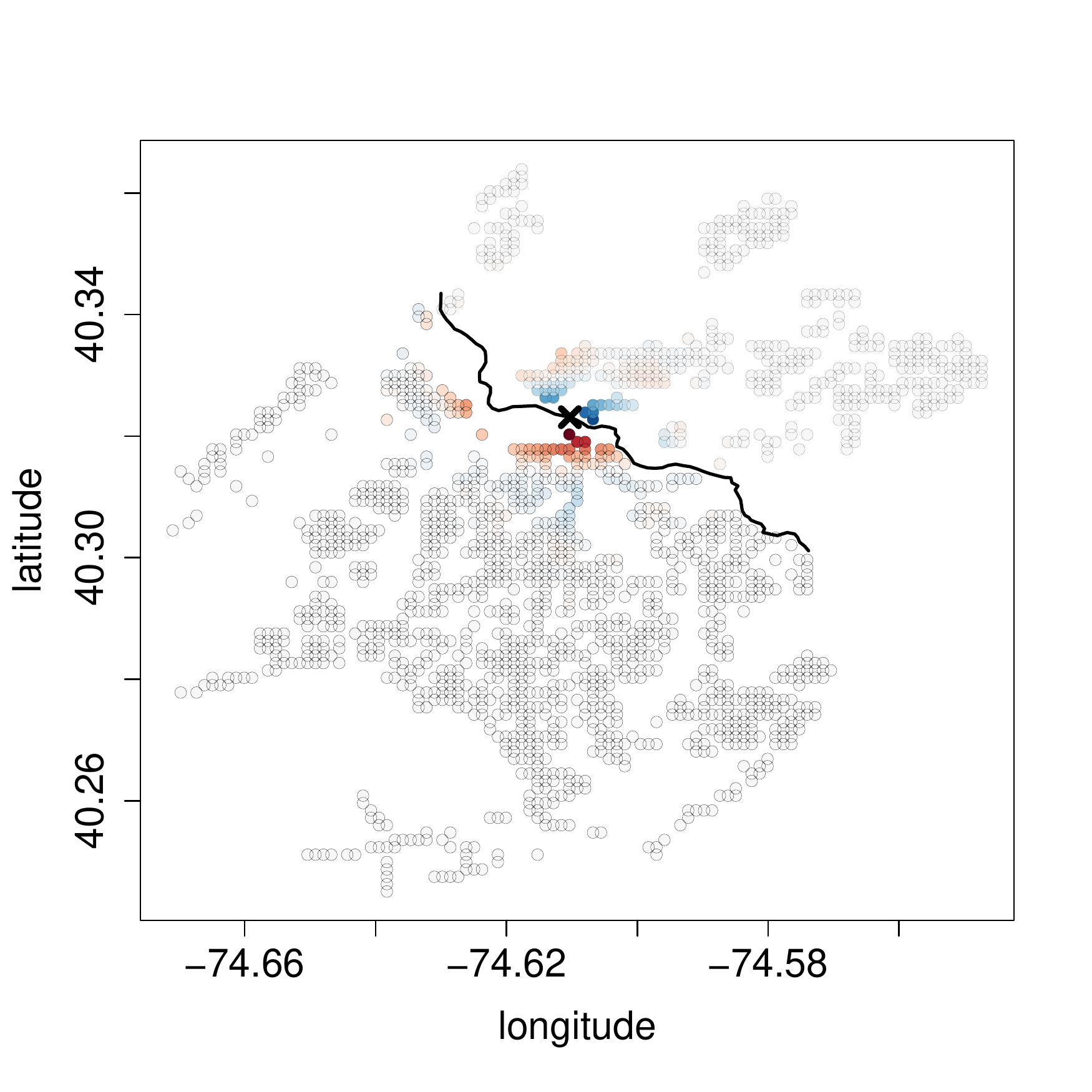} &
\includegraphics[width=0.45\textwidth, trim=7mm 7mm 7mm 7mm]{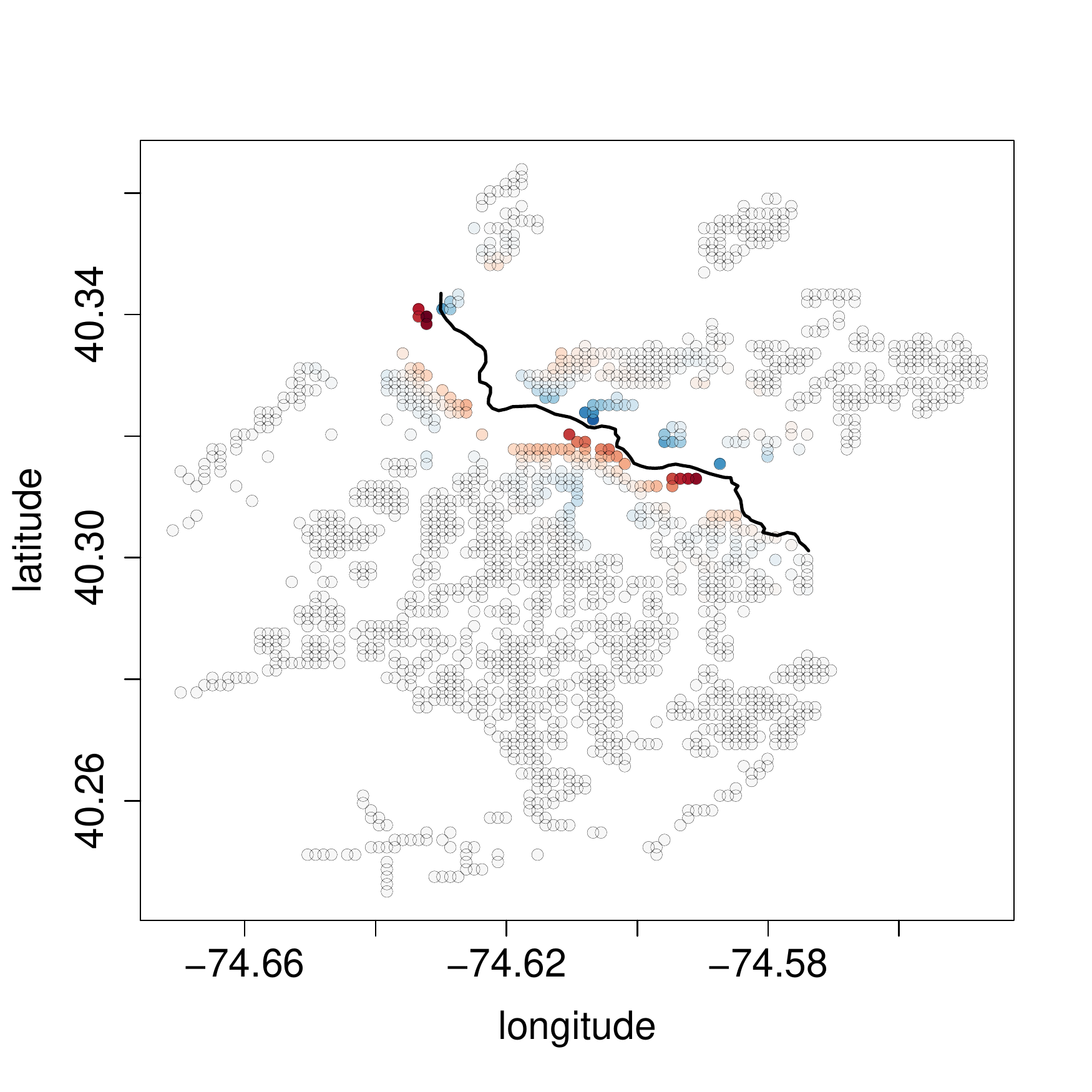}
\end{tabular}
\caption{Weighting functions for the geographic regression discontinuity example of 
\citet{keele2014geographic}, where points depict potential voters within a single school district
and the solid black line is a media market boundary. The left panel depicts an optimal weighting function for
the conditional average treatment effect at the point $c$ marked with a boldface-$\times$ as in \eqref{eq:estimator_2d_pt},
while the right one allows for a weighted treatment effect as in \eqref{eq:estimator_2d_greedy} or,
equivalently, shows the optimal weighting function for a constant effect.
Households below the line are treated (i.e., in the Philadelphia media market), whereas
those above it are controls (i.e., in the New York media market).
The color of the point depicts the $\gamma$-weight: Red points receive positive weight and
blue points receive negative weight, while the shading indicates absolute value of the weight
(darker is larger).}
\label{fig:geo_weights}
\end{figure}

To gain intuition for the multivariate version of our method, we
outline a simple example building on the work of \citet{keele2014geographic}
on the effect of television advertising on voter turnout in presidential elections.
To estimate this effect, \citet{keele2014geographic} examine a school district in New Jersey,
half of which belongs to the Philadelphia media market and
the other half to the New York media market. Before the 2008 presidential elections, the Philadelphia half was
subject to heavy campaign advertising whereas the New York half was not, thus creating a natural experiment
for the effect of television advertising assuming the media market boundary didn't coincide with other major
boundaries within the school district. \citet{keele2014geographic} use this identification strategy to build
a regression discontinuity design, comparing sets of households straddling the media market boundary.

However, despite the multivariate identification strategy, \citet{keele2014geographic} then reduce the
problem to a univariate regression discontinuity problem for estimation: They first compute Euclidean distances
$D_i = \Norm{X_i - c}_2$ to a focal point $c$, and then use $D_i$ as a univariate running variable.
In contrast, our approach allows for transparent inference without needing to rely on such a reduction.
Figure \ref{fig:geo_weights} depicts $\gamma$-weights generated by our optimized approach; the
resulting treatment effect estimator is then $\sum_i \gamma_i Y_i$.
Qualitatively, we replicate the ``no measurable effect'' finding of \citet{keele2014geographic},
while directly and uniformly controlling for spatial curvature effects.
We discuss details, including placebo diagnostics and the choice of tuning parameter,
in Section \ref{sec:keele_titiunik}.

We also see that, at least here, standard heuristics used to reduce the multivariate regression discontinuity problem to a
univariate one are not sharp. In the setup of the left panel of Figure \ref{fig:geo_weights}, where we seek
to estimate the treatment effect at a focal point $c$, some treated points due west of $c$ get a positive weight, whereas
points the same distance south from $c$ get a mildly negative weight, thus violating the heuristic of
\citet{keele2014geographic} that weights should only depend on $D_i = \Norm{X_i - c}_2$. Meanwhile, we can
compare the approach in the right panel of Figure \ref{fig:geo_weights}, where we allow for averaging of treatment
effects along the boundary, to the popular heuristic of using shortest distance to the boundary of the treatment region 
as a univariate running variable \citep{black1999better}. But this reduction again does not capture the behavior of our optimized
estimator: There are some points at the eastern edge of the treated region that are very close to the boundary, but
get essentially zero weight (presumably because there are no nearby units on the control side of the boundary).

\subsection{Related Work}
\label{sec:literature}

The idea of constructing estimators of the type \eqref{eq:estimator} that are minimax with respect
to a regularity class for the underlying data-generating process has a long history in statistics. In early 
work, \citet{legostaeva1971minimax} and \citet{sacks1978linear} independently studied inference
in ``almost'' linear models that arise from taking a Taylor expansion around a point; see also \citet{cheng1997automatic}.
For a broader discussion of minimax linear estimation over non-parametric function classes,
see \citet{cai2003note}, \citet{donoho1994statistical},  \citet{ibragimov1985nonparametric},
\citet{johnstone2011gaussian}, \citet{juditsky2009nonparametric}, and references therein.
An important result in this literature that, for many problems of interest, minimax linear estimators are
within a small explicit constant of being minimax among all estimators \citep{donoho1991geometrizing}.

\citet{armstrong2016optimal} apply these methods to regression discontinuity designs, resulting
in an estimator of the form \eqref{eq:estimator}, except with weights\footnote{\citet{armstrong2016optimal}
also consider a more general setting where we assume accuracy of the $k$-th order Taylor expansion of $\mu_w(x)$ around
$c$; and, in fact, our method also extends to this setting. Here, however, we focus on second-derivative
bounds, which are by far the most common in applications.}
\begin{align}
\label{eq:ak_weights}
&\hgamma = \argmin_{\gamma} \cb{\sum_{i = 1}^n \gamma_i^2 \sigma_i^2 + A^2_B(\gamma)}, \\
\notag
&A_B(\gamma) = \sup_{\mu_0(\cdot),\mu_1(\cdot)} \cb{\sum_{i = 1}^n \gamma_i \mu_{W_i}(X_i) - \tau(c) : \abs{\mu_w(x) - \mu_w(c) - \mu'_w(c) (x - c)} \leq \frac{B}{2} (x - c)^2}.
\end{align}
Now, although this class of functions is cosmetically quite similar to the bounded-second-derivative
class used in \eqref{eq:estimator}, we note that the class of weights allowed for in \eqref{eq:ak_weights} is substantially larger, even if the value of $B$ is the same. This is because the functions $\mu_w(\cdot)$ underlying the above
weighting scheme need not be continuous, and can in fact have jumps of magnitude $B(x - c)^2/2$.
Given that the key assumption underlying regression discontinuity designs is continuity of the conditional means of the potential outcomes at the threshold for the running variable, it would appear to be reasonable to impose continuity away from the threshold as well.
Allowing for jumps through the condition \eqref{eq:ak_weights}
can make the resulting confidence intervals for $\tau(c)$ substantially larger than they are under the smoothness condition with bounded second derivatives.
One key motivation for the weighting scheme \eqref{eq:ak_weights} rather than our proposed
one \eqref{eq:estimator} appears to be that the optimization problem induced
by \eqref{eq:ak_weights} is substantially easier, and allows for closed-form solutions for \smash{$\hgamma_i$}.
Conversely, we are aware of no closed-form solution for \eqref{eq:estimator}, and instead need to rely on
numeric convex optimization.

In the special case where the running variable $X$ is assumed to have a continuous density around the
threshold $c$, there has been a considerable number of recent proposals for asymptotic confidence
intervals while imposing smoothness assumptions on $\mu_w(x)$. \citet*{calonico2014robust} propose a bias correction
to the local linear regression estimator that allows for valid inference, and \citet*{calonico2017effect} provide
further evidence that such bias corrections may be preferable to undersmoothing.
Meanwhile, \citet{armstrong2016simple} show that when $\mu_w(x)$ is twice differentiable
and $X$ has a continuous density around $c$,
we can use local linear regression with a bandwidth chosen to optimize mean-squared error as
the basis for bias-adjusted confidence intervals, provided we inflate confidence intervals by an
appropriate, universal constant (e.g., to build 95\% confidence intervals, one should use a critical
threshold of 2.18 instead of 1.96). \citet{gao2017minimax} characterizes the asymptotically optimal
kernel for the regression discontinuity parameter under the bounded second derivative assumption
with a continuous running variable.
As discussed above, the value of our approach relative to this literature is that we allow for considerably
more generality in the specification of the regression discontinuity design: the running
variable $X$ may be discrete and/or multivariate, and the treatment boundary may be irregularly shaped.

Optimal inference with multiple running variables is less developed than in the univariate case. \citet{papay2011extending} and
\citet{reardon2012regression} study local linear regression with a ``small'' bandwidth,
but do not account for finite sample bias due to curvature. \citet{zajonc2012regression}
extends the analysis of \citet{imbens2011optimal} to the multivariate case, and studies optimal bandwidth
selection for continuous running variables given second derivative bounds; the inference,
however, again requires undersmoothing. \citet*{keele2015enhancing} consider an approach to geographic
regression discontinuity designs based on matching. To our knowledge, the approach we present below 
is the first to allow for uniform, bias-adjusted inference in the multivariate regression discontinuity setting.

Finally, although local methods for inference in the regression discontinuity design have desirable theoretical
properties, many practitioners also seek to estimate $\tau(c)$ by fitting
$\EE{Y_i \cond X_i = x}$ using a global polynomial expansion along with a jump at $c$; see \citet{lee2010regression}
for a review and examples. However, as argued by
\citet{gelman2014high}, this approach is not recommended, as the model with the best in-sample fit may provide
poor estimates of the discontinuity parameter. For example, high-order polynomials may give
large influence to samples $i$ for which $X_i$ is far from the decision boundary $c$, and thus lead
to unreliable performance.

Another approach to regression discontinuity designs (including in the discrete case) builds on randomization inference;
see \citet*{cattaneo2015randomization}, \citet*{cattaneo2017practical} and \citet*{li2015evaluating}
for a discussion. The problem of specification testing for regression discontinuity designs is considered by
\citet*{cattaneo2016simple}, \citet{frandsen2017party} and \citet{mccrary2008manipulation}.

\section{Formal Considerations}
\label{sec:theory}

\subsection{Uniform Asymptotic Inference}
\label{sec:inference}

Our main result verifies that optimized designs can be used for valid asymptotic inference
about $\tau(c)$. We here consider the problem of estimating a conditional average treatment
effect at a point $c$ as in \eqref{eq:estimator} or \eqref{eq:estimator_2d_pt}; similar arguments
extend directly to the averaging case as in \eqref{eq:estimator_2d_greedy}.
Following, e.g., \citet{robins2006adaptive}, we seek confidence intervals $\ii_\alpha$
that attain uniform coverage over the whole regularity set under consideration:
\begin{equation}
\label{eq:uniform}
\liminf_{n \rightarrow \infty} \ \inf\cb{\PP{\mu_1(c) - \mu_0(c) \in \ii_\alpha} : \Norm{\nabla^2 \mu_{w}(x)} \leq B \text{ for all } w, \, x} \geq 1 - \alpha.
\end{equation}
As in \citet{armstrong2016optimal}, our approach to building such confidence intervals
relies on an explicit characterization of the bias of $\htau$ rather than on undersmoothing.
Our key result is as follows.

\begin{theo}
\label{theo:gauss}
Suppose that we have a moment bound $\mathbb{E}[(Y_i - \EE{Y_i \cond X_i})^q \cond X_i = x] \leq C$
uniformly over all $x \in \RR^k$, for some exponent $q > 2$ and constant $C \geq 0$. Suppose,
moreover, that $0 < \sigma_{\min} \leq \sigma_i$ for all $i = 1, \, ..., \, n$ for a deterministic
value $\sigma_{\min}$, and that none of the weights $\hgamma_i$ derived in \eqref{eq:estimator} or \eqref{eq:estimator_2d_pt}
dominates all the others, i.e.,\footnote{The bound on the relative contribution of any single \smash{$\hgamma_i$}
is needed to obtain a Gaussian limit distribution for \smash{$\htau$}. In related literature,
\citet{armstrong2016optimal} follow \citet{donoho1994statistical}, and side-step this issue by assuming
Gaussian errors $Y_i(w) - \mu_w(X_i)$, in which case no central limit theorem is needed.
Conversely, \citet*{athey2016efficient} adopt an approach more similar to ours, and explicitly bound
$\hgamma_i$ from above during the optimization.}
\begin{equation}
\label{eq:gamma_sup}
\max_{1 \leq i \leq n}\cb{\hgamma_i^2} \, \Big/ \, \sum_{i = 1}^n \hgamma_i^2 \rightarrow_p 0.
\end{equation}
Then, our estimator $\htau$ from \eqref{eq:estimator} or \eqref{eq:estimator_2d_pt} is asymptotically Gaussian,
\begin{equation}
\label{eq:gauss}
\p{\htau - b\p{\hgamma}} \,\big/\, s\p{\hgamma} \Rightarrow \nn\p{0, \, 1},  \ \ b(\hgamma) = \sum_{i = 1}^n \hgamma_i \mu_{W_i}(X_i) - \tau(c), \ \ s^2\p{\hgamma} := \sum_{i = 1}^n \hgamma_i^2 \sigma_i^2,
\end{equation}
where $b\p{\hgamma}$ denotes the conditional bias,
and $s^2\p{\hgamma} \rightarrow_p 0$.
\end{theo}

Now, in solving the optimization problem \eqref{eq:estimator}, we also obtain an explicit bound $\hat{t}$
on the conditional bias, \smash{$b(\hgamma) \leq \hat{t}$}, and so can use the following
natural construction to obtain confidence intervals for $\tau(c)$ \citep{imbens2004confidence},
\begin{equation}
\label{eq:CI}
\tau(c) \in \htau \pm l_\alpha, \ \ l_\alpha = \min\cb{l : \PP{\abs{b + s\p{\hgamma} Z} \leq l} \geq \alpha \text{ for all } \abs{b} \leq \hat{t}}, \ \ Z \sim \nn\p{0, \, 1},
\end{equation}
where $\alpha$ is the significance level. These confidence intervals are asymptotically
uniformly valid in the sense of \eqref{eq:uniform}, i.e., for any $\alpha' < \alpha$, there is a threshold $n_{\alpha'}$
for which, if $n \geq n_{\alpha'}$, the confidence intervals \eqref{eq:CI} achieve $\alpha'$-level coverage for any functions
$\mu_{w}(\cdot)$ in our regularity class.

Finally, whenever $X_i$ does not have support arbitrarily close to $c$, e.g., in the case where $X_i$ has
a discrete distribution, the parameter $\tau(c)$ is not point identified. Rather, even with infinite data, the
strongest statement we could make is that
\begin{equation}
\begin{split}
&\tau(c) \in \ii^*, \ \ \ii^* = \text{range}\Big\{\mu_{(1)}(c) - \mu_{(0)}(c) : \Norm{\nabla^2 \mu_{(w)}(x)} \leq B, \eqand \\
&\ \ \ \ \mu_{(w)}(x) = \EE{Y_i \cond X_i = x, \, W_i = w} 
\text{ for all } x, \, w \in \text{supp}\cb{X_i, \, W_i}, \Big \},
\end{split}
\end{equation}
where $\text{supp}\cb{X_i, \, W_i}$ denotes the support of $(X_i, \, W_i)$. In this case, our confidence intervals
\eqref{eq:CI} remain valid for $\tau(c)$; however, they may not cover the whole
optimal identification interval $\ii^*$. In partially identified settings, this type of confidence intervals
(i.e., ones that cover the parameter of interest but not necessarily the whole identification interval)
are advocated by \citet{imbens2004confidence}. From the perspective of the practitioner, an advantage of
our approach is that intervals for $\tau(c)$ have the same interpretation whether or not $\tau(c)$ is point identified,
i.e., uniformly in large samples, $\tau(c)$ will be covered with probability $1 - \alpha$. Then, asymptotically,
intervals \eqref{eq:CI} will converge to a point if and only if $\tau(c)$ is point identified. For a further discussion of
regression discontinuity inference with discrete running variables, see \citet{kolesar2016inference}.

\subsection{Implementation via Convex Optimization}
\label{sec:implementation}

In our presentation so far, we have discussed several non-parametric convex optimization problems,
and argued that solving them was feasible given advances in the numerical optimization literature
over the past few decades \citep[e.g.,][]{boyd2004convex}. Here, we present a concrete solution strategy via quadratic
programming over a discrete grid, and show that the resulting discrete solutions converge uniformly to the continuous
solution as the grid size becomes small.\footnote{In the case where $X$ is univariate, the resulting optimization
problem is a classical one, and arguments made by \citet{karlin1973some} imply that the weights $\hgamma_i$
can be written as $\hgamma_i = g(X_i)$ where $g$ is a perfect spline; and our proposed optimization strategy
reflects this fact. However, in the multivariate case, we are not aware of a similar simple characterization.}

To do so, we start by writing the optimization problems underlying \eqref{eq:estimator}, \eqref{eq:estimator_2d_pt}
and \eqref{eq:estimator_2d_greedy} in a unified form. Given a specified focal point $c$, we seek to solve
\begin{equation}
\label{eq:qp_general}
\begin{split}
&\underset{\gamma, \, t}{\text{minimize}} \ \sum_{i = 1}^n \gamma_i^2 \sigma_i^2 + B^2 t^2 \text{ subject to }\\
&\ \ \ \  \sum_{i = 1}^n \gamma_i \p{f_0(X_i) + \psi \, w(X_i) \p{f_1(X_i) - f_0(X_i)}}  \leq t, \\
&\ \ \ \ \ \ \ \ \ \ \text{for all } f_w(c) = 0, \ \nabla f_w(c) = 0, \ \Norm{\nabla^2 f_w(x)} \leq 1 \text{ with }  w \in \cb{0, \, 1}\\
&\ \ \ \ \sum_{i = 1}^n w(X_i) \gamma_i = 1, \ \ \sum_{i = 1}^n (1 - w(X_i)) \gamma_i = -1, \\ 
&\ \ \ \ \sum_{i = 1}^n \gamma_i (X_i - c)= 0, \ \ \psi \sum_{i = 1}^n \p{2w(X_i) - 1} \gamma_i (X_i - c)= 0,
\end{split}
\end{equation}
where $w(x)$ denotes the treatment assignment function, and
$\psi$ lets us toggle between different problem types.
If we want to estimate the CATE at $c$ as in \eqref{eq:estimator_2d_pt} we set $\psi = 1$,
whereas if we want an optimally weighted CATE estimator as in \eqref{eq:estimator_2d_greedy} we set $\psi = 0$.

To further characterize the solution to this problem, we can use Slater's constraint qualification
\citep[e.g.,][Theorem 3.11.2]{ponstein2004approaches} to verify that strong duality holds,
and that the optimum of \eqref{eq:qp_general} matches the optimum of the following problem:
\begin{equation}
\label{eq:dual}
\begin{split}
&\underset{f(\cdot), \, \lambda}{\text{maximize}} \ \inf_{\gamma, \, t} \Bigg\{\sum_{i = 1}^n \gamma_i^2 \sigma_i^2 +B^2  t^2 \\
&\ \ \ \ \ \
+ \lambda_1 \p{\sum_{i = 1}^n \gamma_i \p{f_0(X_i) + \psi \, w(X_i) \p{f_1(X_i) - f_0(X_i)}}  - t} \\
&\ \ \ \ \ \
+ \lambda_2 \p{\sum_{i = 1}^n \gamma_i  w(X_i) - 1} 
+ \lambda_3 \p{\sum_{i = 1}^n \gamma_i (1 - w(X_i)) + 1} \\
&\ \ \ \ \ \
+  \sum_{i = 1}^n \gamma_i \p{\lambda_4  + \psi \, \lambda_5 (2w(X_i) - 1)} (X_i - c) \Bigg\} \\
&\text{subject to } f_w(c) = 0, \ \nabla f_w(c) = 0, \ \Norm{\nabla^2 f_w(x)} \leq 1 \text{ for } w \in \cb{0, \, 1}, \\
&\ \ \ \ \ \
\lambda_1, \geq 0, \ \lambda_2, \, \lambda_3 \in \RR, \ \lambda_4, \, \lambda_5 \in \RR^k,
\end{split}
\end{equation}
where $k$ is the number of running variables.
Here, we also implicitly used von Neumann's minimax theorem to move
the maximization over $f$ outside the $\inf_{\gamma, \, t}\{\}$ statement.

The advantage of this dual representation is that, by examining first order conditions in the $\inf_{\gamma, \, t}\{\}$
term, we can analytically solve for $\gamma$ and $t$ in the dual objective, e.g.,
\begin{equation}
-2 \sigma_i^2 \hgamma_i = \hlambda_1 \p{\hf_0(X_i) + \psi \, w(X_i) \p{\hf_1(X_i) - \hf_0(X_i)}}
+ \hlambda_2 w(X_i)  +  \ldots,
\end{equation}
where $\hf_0(\cdot)$, $\hf_1(\cdot)$, $\hlambda_1$, etc., are the maximizers of \eqref{eq:dual}.
Carrying out the substitution results in a more tractable optimization problem over the space of twice
differentiable functions $f$, along with a finite number of Lagrange parameters $\lambda_j$:
\begin{equation}
\label{eq:dual_simplified}
\begin{split}
&\underset{\tf(\cdot), \, \lambda}{\text{minimize}} \ \frac{1}{4} \sum_{i = 1}^n \frac{G_i^2}{\sigma_i^{2}} + \frac{1}{4} \frac{\lambda_1^2}{B^2} + \lambda_2 - \lambda_3 \\
&\text{subject to } G_i = \tf(X_i) + \lambda_2 w(X_i) + \lambda_3 (1 - w(X_i))  \\
&\ \ \ \ \ \ \ \ \ \ \ \ \ \ \ \ \ \ \ \ \ \ \ \ \ \ \ \ + \lambda_4 (X_i - c) + \psi \, \lambda_5 (2w(X_i) - 1) (X_i - c) \\
&\ \ \tf(x) = \tf_0(x) + \psi \, w(x) \p{\tf_1(x) - \tf_0(x)}, \ \lambda_1 \geq 0, \ \lambda_2, \, \lambda_3 \in \RR, \ \lambda_4, \, \lambda_5 \in \RR^k, \\
&\ \ \tf_w(c) = 0, \ \nabla\tf_w(c) = 0, \ \Norm{\nabla^2 \tf_w(x)} \leq \lambda_1 \text{ for } w \in \cb{0, \, 1},
\end{split}
\end{equation}
where $\tf_w(x)$ in the above problem corresponds to $\lambda_1 f_w(x)$ in \eqref{eq:dual},
and we can recover our weights of interest via \smash{$\hgamma_i = -\sigma_i^{-2} \hG_i /2$}
and \smash{$\hat{t} = \hlambda_1 / (2B^2)$}.
The upshot of these manipulations is that \eqref{eq:dual_simplified} can be approximated via a
finite-dimensional quadratic program. In our software implementation \texttt{optrdd}, we use this type of a finite dimensional
approximation to obtain \smash{$\hgamma_i$} following the construction described in the proof
of Proposition \ref{prop:approx}.

\begin{prop}
\label{prop:approx}
Suppose that $X_i \in \xx$ belong to some compact, convex set $\xx \subset \RR^k$. Then,
for any tolerance level $\eta > 0$, there exists a finite-dimensional quadratic program that can recover
the solution \smash{$\hgamma$} to \eqref{eq:dual_simplified} with $L_2$-error at most $\eta$.
\end{prop}

\subsection{Minimizing Confidence Interval Length}

As formulated in \eqref{eq:estimator}, our estimator seeks to minimize the worst-case
mean-squared error over the specified bounded-second-derivative class. However, in some
applications, we may be more interested in making the confidence intervals \eqref{eq:CI} as
short as possible; and our approach can easily be adapted for this objective.
To do so, consider the minimization objective in \eqref{eq:qp_general}.
Writing \smash{$\hat{v}^2 = \sum_{i = 1}^n \hgamma_i^2 \sigma_i^2$}, we see that
both the worst-case mean-squared error, \smash{$\hat{v}^2 + B^2 \hat{t}^2$}, and the confidence interval
length in \eqref{eq:CI} are monotone increasing functions of \smash{$\hat{v}$} and \smash{$\hat{t}$}; the
only difference is in how they weight these two quantities at the optimum.

Now, to derive the full Pareto frontier of pairs \smash{$(\hat{v}, \, \hat{t})$}, we can simply re-run \eqref{eq:qp_general}
with the term $B^2t^2$ in the objective replaced with $\lambda B^2t^2$, for some $\lambda > 0$.
A practitioner wanting to minimize the length of confidence intervals could consider computing this whole
solution path to \eqref{eq:qp_general}, and then using the value of $\lambda$ that yields the best intervals;
this construction provides minimax linear fixed-length confidence intervals \citep{donoho1994statistical}.
Since this procedure never looks at the responses $Y_i$, the inferential guarantees for
the resulting confidence intervals remain valid.

In our applications, however, we did not find a meaningful gain from optimizing over $\lambda$ instead
of just minimizing worst-case mean-squared error as in \eqref{eq:qp_general}, and so did not pursue this line of investigation
further. This observation is in line with the analytic results of \citet{armstrong2016simple} who showed that,
when $X$ has a continuous density and $\mu_w(x)$ is twice differentiable, using the mean-squared error optimal
bandwidth for local linear regression is over 99\% efficient relative to using a bandwidth that minimizes the length of
bias-adjusted confidence intervals.

Finally, although it is beyond the scope of the present paper, it is interesting to ask whether
we can generalize our approach to obtain asymptotically quasi-minimax estimators for $\tau(c)$
when the per-observation noise-scale $\sigma_i$ needs to be estimated from the data.
The resulting question is closely related to the classical issue of when feasible generalized least
squares can emulate generalized least squares; see \citet{romano2017resurrecting} for a recent
discussion.

\section{Univariate Optimized Designs in Practice}
\label{sec:tuning}

To use this result in practice, we of course need to estimate the sum \smash{$\sum \hgamma_i^2 \sigma_i^2$} and
choose a bound $B$ on curvature. Estimating the former is relatively routine; and we recommend the following.
First, we estimate $\mu_w(x)$ globally, or over a large plausible relevant interval around the threshold,
and average the square of the residuals \smash{$R_i = Y_i - \hmu_{W_i}(X_i)$} to obtain an estimate \smash{$\hsigma^2$}
of the average value of $\sigma_i^2$. Then, we optimize weights \smash{$\hgamma_i$} using \eqref{eq:estimator},
with \smash{$\sigma_i^2 \leftarrow \hsigma^2$}. Finally, once we have chosen the weights $\gamma_i$, we estimate the
sampling error of \smash{$\htau$} as below, noting that the estimator will be consistent under standard conditions
\begin{equation}
\label{eq:sigma_hat}
\hs^2\p{\hgamma} = \sum_{i = 1}^n \hgamma_i^2 \p{Y_i - \hmu_{W_i}(X_i)}^2, \ \
\hs^2\p{\hgamma} \, \big/\, \sum \hgamma_i^2 \sigma_i^2 \geq 1 - o_P(1).
\end{equation}
Conceptually, this strategy is comparable to first running local linear regression without heteroskedasticity
adjustments to get a point estimate, but then ensuring that the uncertainty quantification is
heteroskedasticity-robust \citep{white1980heteroskedasticity}.
We summarize the resulting method as Procedure \ref{alg:optrdd}.
We always encourage plotting the weights \smash{$\hgamma_i$}
against $X_i$ when applying our method.

Conversely, obtaining good bounds on the curvature $B$ is more difficult,
and requires problem specific insight. In particular, adapting to the true curvature
$\mu_{w}(x)$ without a-priori bounds for $B$ is not always possible; see
\citet{armstrong2016optimal}, \citet{bertanha2016impossible}, and references therein. In applications, we recommend
considering a range of plausible values of $B$ that could be obtained, e.g., from subject-matter
expertise or from considering the mean-response function globally. For example,
we could estimate $\mu_w(x)$ using a quadratic function globally, or over a large plausible relevant
interval around the threshold, and then multiply maximal curvature of the fitted model by
a constant, e.g., 2 or 4. The larger the value of $B$ we use the more conservative the resulting
inference. In practice, it is often helpful to conduct a sensitivity analysis for the robustness of confidence
intervals to changing $B$; see Figure \ref{fig:supplem_fig} for an example.\footnote{An interesting wrinkle is that
if we are able to bound $B$ in large samples---but not uniformly---then confidence intervals built using estimated
values of $B$ will have asymptotic but not uniform coverage.}

\begin{algbox}[t]
\myalg{alg:optrdd}{Optimized Regression Discontinuity Inference}{
This algorithm provides confidence intervals for the conditional average
treatment effect $\tau(c)$, given an a-priori bound $B$ on the second derivative
of the functions $\mu_w(x)$. We assume that the conditional variance parameters
$\sigma_i^2$ are unknown; if they are known, they should be used as in \eqref{eq:estimator}.
This procedure is implemented in our \texttt{R} package \texttt{optrdd}.\footnotemark{}
\begin{enumerate}
\item Pick a large window $r$, such that data with $\abs{X_i - c} > r$ can be safely
ignored without loss of efficiency. (Here, we can select $r = \infty$, but this may result in
unnecessary computational burden.)
\item Run ordinary least-squares regression of $Y_i$ on the interaction of $X_i$ and $W_i$ over the
window $\abs{X_i - c} \leq r$. Let \smash{$\hsigma^2$} be the residual error from this regression.
\item Obtain $\hgamma$ via the quadratic program \eqref{eq:qp_general}, with $\sigma_i$ set
to $\hsigma$ and weights outside of the range $\abs{X_i - c} \leq r$ set to 0.
\item Confirm that the optimized weights $\hgamma_i$ are small for $\abs{X_i - c} \approx r$.
If not, start again with a larger value of $r$.\footnotemark{}
\item Estimate \smash{$\htau = \sum_{i = 1}^n \hgamma_i Y_i$} and
\smash{$\hs^2 = \sum_{i = 1}^n \hgamma_i^2 (Y_i - \hmu_{W_i}(X_i))^2$}, where
the \smash{$\hmu_{W_i}(X_i)$} are predictions from the least squares regression
from step 1.
\item Build confidence intervals as in \eqref{eq:CI}.
\end{enumerate}
}
\end{algbox}

\footnotetext[11]{Here, the algorithm assumes that all observations are of roughly the same quality
(i.e., we do not know that $\sigma_i^2$ is lower for some observations than others). If we have
a-priori information about the relative magnitudes of the conditional variances of different observations,
e.g., some pairs outcomes $Y_i$ are actually aggregated over many observations, then we should run
steps 2 and 3 below using appropriate inverse-variance weights. Our software allows for such weighting.}

\footnotetext{Only considering data over an a-priori specified ``large plausible relevant interval'' around $c$
that safely contains all the data relevant to fitting $\tau(c)$ can also be of computational interest.
Our method relies on estimating a smooth non-parametric function over the whole range
of $x$; and being able to reduce the relevant range of $x$ a-priori reduces the required computation.
Although defining such plausibility intervals is of course heuristic, our method
ought not be too sensitive to how exactly the interval was chosen. For example, in the setup considered
in Section \ref{sec:oreopoulos}, the optimal bandwidth for local linear regression is around 3 or 6 years depending
on the amount of assumed smoothness (and choosing a good bandwidth is very important); conversely, using
plausibility intervals extending 10, 15, or 20 years on both sides of $c$ appears to work reasonably well. 
When running the method \eqref{eq:estimator}, one should always make sure that the weights
\smash{$\hgamma_i$} get very small near the edge of the plausibility interval; if not, the interval should be
made larger.}

\subsection{Application: The Effect of Compulsory Schooling}
\label{sec:oreopoulos}

In our first application, we consider a dataset from \citet{oreopoulos2006estimating}, who studied the
effect of raising the minimum school-leaving age on earnings as an adult. The effect is identified by the
UK changing its minimum school-leaving age from 14 to 15 in 1947, and the response is log-earnings
among those with non-zero earnings (in 1998 pounds).
This dataset exhibits notable discreteness in its running variable, and was used by \citet{kolesar2016inference}
to illustrate the value of their bias-adjusted confidence intervals for discrete regression discontinuity designs.
For our analysis, we pre-process our data exactly as in \citet{kolesar2016inference}; we refer the reader to their
paper and to \citet{oreopoulos2006estimating} for a more in-depth discussion of the data.

As in \citet{kolesar2016inference}, we seek to identify the effect of the change in minimum school-leaving age
on average earnings via a local analysis around the regression discontinuity; our running variable is the year in
which a person turned 14, with a treatment threshold at 1947.
\citet{kolesar2016inference} consider analysis using local linear regression with a rectangular kernel and a bandwidth chosen
such as to make to make their honest confidence intervals as short as possible (recall that we can measure confidence interval
length without knowing the point estimate, and so tuning the interval length does not invalidate inference). Here,
we also consider local linear regression with a triangular kernel, as well as our optimized design.\footnote{ \citet{oreopoulos2006estimating}
analyze the dataset using a global polynomial specification with clustered random variables, following \citet{lee2008regression}.
However, as discussed in detail by \citet{kolesar2016inference}, this approach does not yield valid confidence intervals.}

\begin{figure}[t]
\centering
\includegraphics[width=0.5\textwidth, trim=7mm 7mm 7mm 7mm]{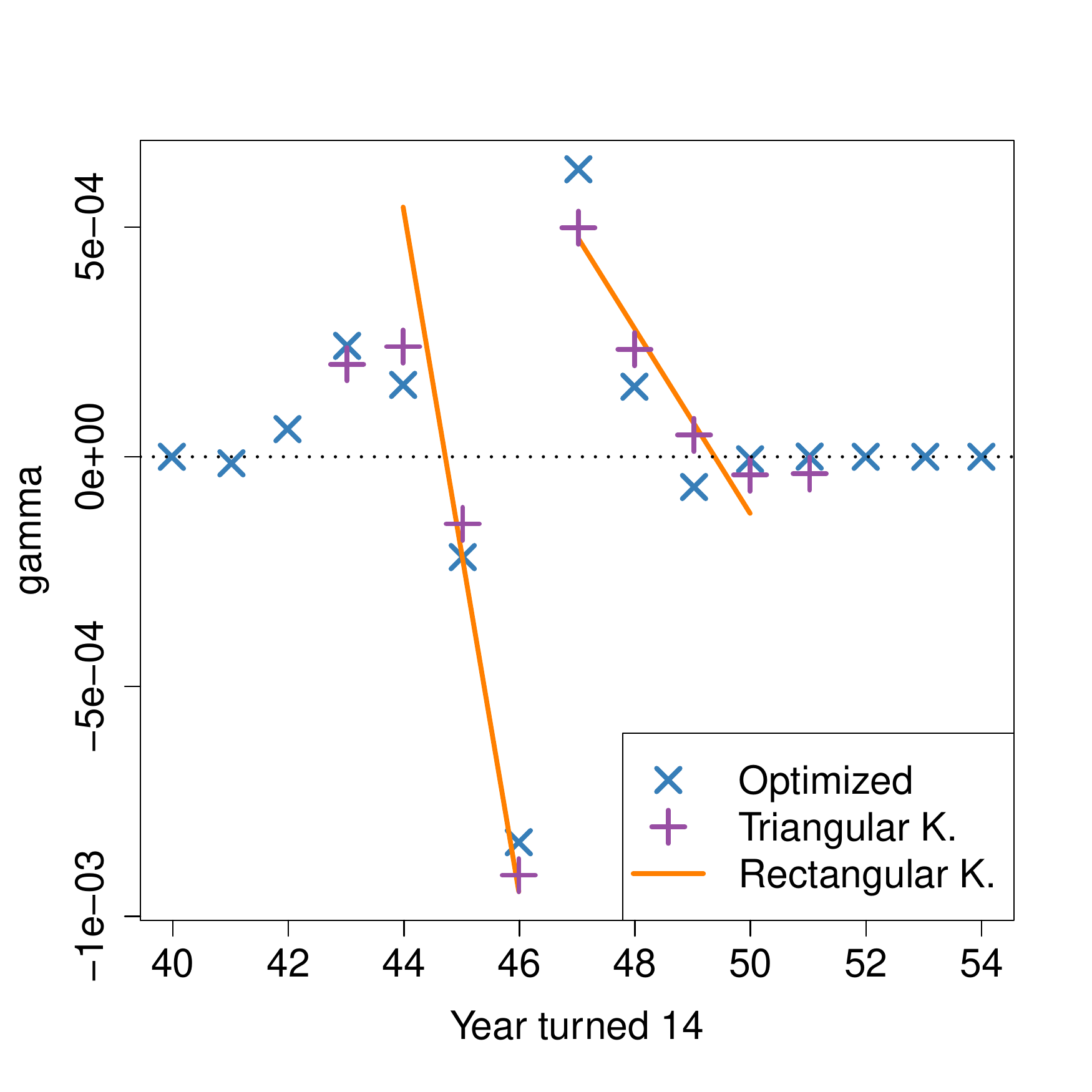}
\caption{Weighting functions \smash{$\hgamma(X_i)$} produced explicitly by our estimator
\eqref{eq:estimator}, and implicitly via local linear regression with a rectangular or triangular
kernel. Both local linear regression methods have a finite bandwidth, and the effective weights of
\smash{$\hgamma(X_i) = 0$} outside this bandwidth are not shown. The weighting functions
were generated with $B = 0.012$.}
\label{fig:orepoulos_gamma}
\end{figure}

In order to obtain confidence intervals, it remains to choose a bound $B$.
Following the discussion in Section \ref{sec:tuning}, a 2nd-order polynomial fit with a ``large'' bandwidth
of either 12 or 18 has a curvature of $0.006$ (the estimate is insensitive to the choice of large bandwidth);
thus, we tried $B = 0.006$ and $B = 0.012$. We also consider the more extreme choices
$B = 0.003$ and $B = 0.03$. For $\sigma_i^2$, we proceed as discussed
in Section \ref{sec:tuning}. Figure \ref{fig:orepoulos_gamma} shows the effective \smash{$\hgamma(X_i)$}
weighting functions for all 3 considered methods, with $B = 0.012$.

\begin{table}[t]
\centering
\begin{tabular}{|r|ccc|}
\hline
$B$ & rect.~kernel & tri.~kernel & optimized \\ \hline
0.003 & $0.0213 \pm 0.0761$ & $0.0321 \pm 0.0737$ & $0.0302 \pm 0.0716$ \\
0.006 & $0.0578 \pm 0.0894$ & $0.0497 \pm 0.0867$ & $0.0421 \pm 0.0841$ \\
0.012 & $0.0645 \pm 0.1085$ & $0.0633 \pm 0.1037$ & $0.0557 \pm 0.1003$ \\
0.03 & $0.0645 \pm 0.1460$ & $0.0710 \pm 0.1367$ & $0.0710 \pm 0.1329$ \\ \hline
\end{tabular}
\caption{Confidence interval ($\alpha = 95\%$) for the effect of raising the minimum school-leaving
age on average log-earnings, as given by local linear regression with a rectangular kernel, local linear
regression with a triangular kernel, and our optimized method \eqref{eq:estimator}. The confidence intervals
account for curvature effects, provided the second derivative is bounded by $B$.}
\label{tab:oreopoulos}
\end{table}

We present results in Table \ref{tab:oreopoulos}. Overall, these results are in line with those presented in Figure \ref{fig:mse_cmp}.
The optimized method yields materially shorter confidence intervals than local linear regression with a rectangular
kernel: for example, with $B = 0.03$, the rectangular kernel intervals are 11\% longer. In comparison, the triangular
kernel comes closer to matching the performance of our method, although the optimized method still has shorter
confidence intervals. Moreover, when considering comparisons with the triangular kernel, we note that the rectangular
kernel is far more prevalent in practice, and that the motivation for using the triangular kernel often builds on
the optimality results of \citet{cheng1997automatic}. And, once one has set out on a quest for optimal
weighting functions, there appears to be little reason to not just use the actually optimal weighting function
\eqref{eq:estimator}.


Finally, we note that a bound $B$
on the second derivative also implies that the quadratic approximation \eqref{eq:ak_weights} holds with
the same bound $B$. Thus, we could in principle also use the method of \citet{armstrong2016optimal}
to obtain uniform asymptotic confidence intervals here. However, the constraint \eqref{eq:ak_weights}
is weaker than the actual assumption we were willing to make (i.e., that the functions $\mu_w(\cdot)$
have a bounded second derivative), and so the resulting confidence intervals are substantially larger.
Using their approach on this dataset gives confidence intervals of $0.0518 \pm 0.0969$ with $B = 0.006$ and
$0.0682 \pm 0.1760$ with $B = 0.03$; these intervals are not only noticeably longer than our intervals,
but are also longer than the best uniform confidence
intervals we can get using local linear regression with a rectangular kernel as in \citet{kolesar2016inference}.
Thus, the use of numerical convex optimization tools that let us solve \eqref{eq:estimator} instead of
\eqref{eq:ak_weights} can be of considerable value in practice.

\section{Applications with Multivariate Discontinuities}
\label{sec:multi}

\subsection{A Discontinuity Design with Two Cutoffs}
\label{sec:matsudaira}

We next consider the behavior of our method in a specific variant of the multiple running
variable problem motivated by a common inference strategy in education.
Some school districts mandate students to attend summer school if they fail a year-end 
standardized test in either math or reading \citep{jacob2004remedial,matsudaira2008mandatory},
and it is of course important to understand the value of such summer schools. The fact
that students are mandated to summer school based on a sharp test score cutoff suggests
a natural identification strategy via regression discontinuities; however, standard univariate
techniques cannot directly be applied as the regression discontinuity now no longer
occurs along a point, but rather along a surface in the bivariate space encoding
both a student's math and reading scores.

\begin{table}
\centering
\begin{tabular}{|r|cc|cc|}
\hline
& \multicolumn{2}{c|}{fail reading} &  \multicolumn{2}{c|}{pass reading} \\
& fail math & pass math & fail math & pass math \\ \hline
number of students & 3,586 & 1,488 & 10,331 & 15,336 \\
summer school attendance & 69.6\% & 61\% & 52.9\% & 10.6\% \\ \hline
\end{tabular}
\caption{Summary statistics for a subset of the dataset of \citet{matsudaira2008mandatory}.}
\label{tab:matsudaira}
\end{table}

We illustrate our approach using the dataset of \citet{matsudaira2008mandatory}. As discussed above,
the goal is to study the impact of summer school on future test scores, and the effect of summer school is identified by
a regression discontinuity: At the end of the school year, students need to take year-end tests
in math and reading; then, students failing either of these tests are mandated to attend summer
school. Here, we focus on the 2001 class of graduating 5th graders, and filter the sample
to only include the $n = 30,741$ students whose 5th-grade math and reading
scores both fall between 40 points of the passing threshold; this represents $44.7\%$ of the full sample.
\citet{matsudaira2008mandatory} analyzed this dataset with univariate methods,
by using reading score as a running variable and only consider the subset of students who passed
the math exam, etc. This allows for a simple analysis, but may also result in a loss of
precision.\footnote{In order to make a formal power comparison, we need to compare two estimators that
target the same estimand. In the simplest case where $\tau(x) = \tau$ is constant, our estimator
\eqref{eq:estimator_2d_greedy} presents an unambiguous gain in power over those considered
in \citet{matsudaira2008mandatory}.}

We present some summary statistics in Table \ref{tab:matsudaira}. Clearly, not all students mandated
to attend summer school in fact attend, and some students who pass both tests still need to attend for
reasons discussed in \citet{matsudaira2008mandatory}. That being said, the effect of passing tests on summer
school attendance is quite strong and, furthermore, the treatment effect of being mandated to summer school
is interesting in its own right, so here we perform an ``intent to treat'' analysis without considering non-compliance.

\begin{figure}
\centering
\begin{tabular}{cc}
\includegraphics[width=0.45\textwidth]{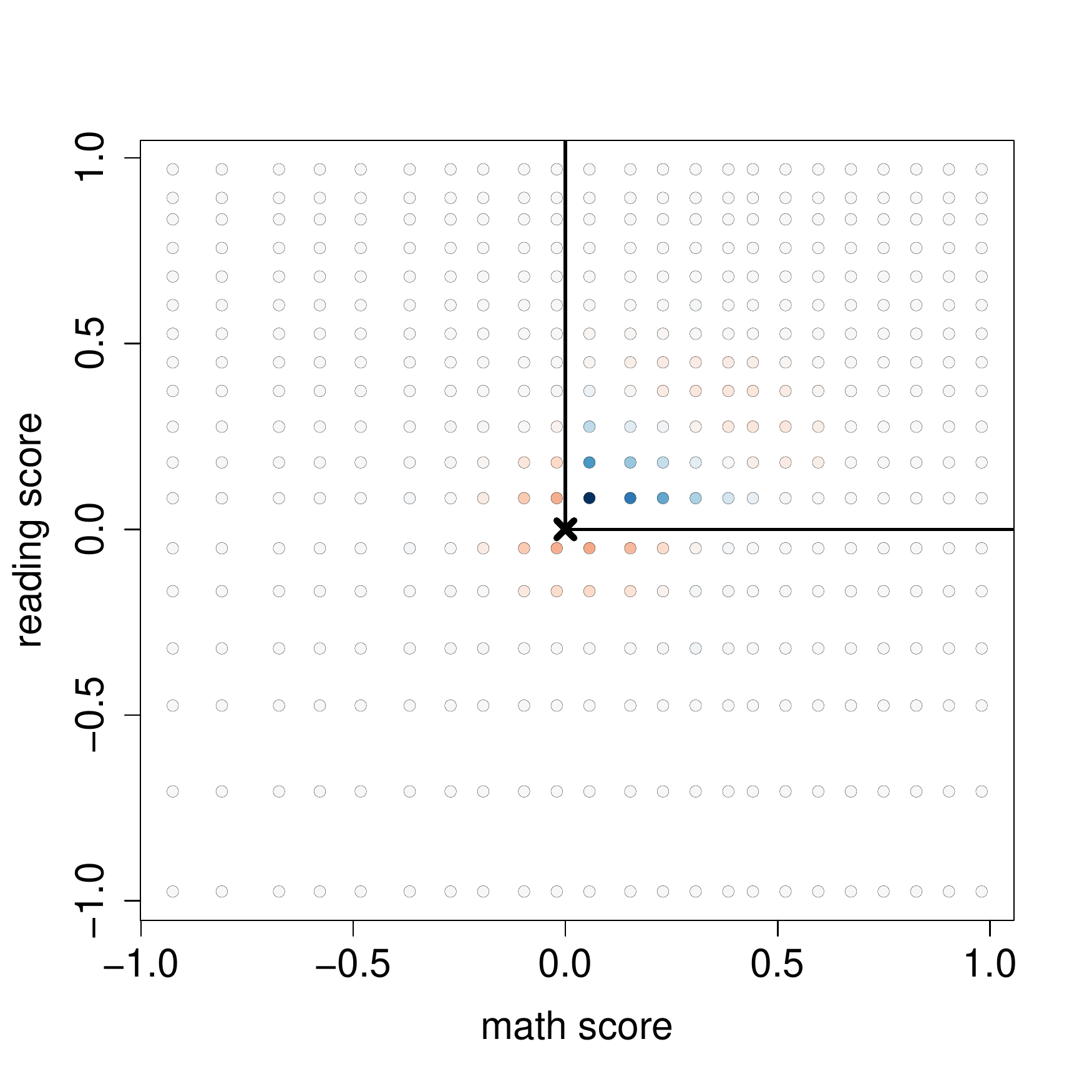} &
\includegraphics[width=0.45\textwidth]{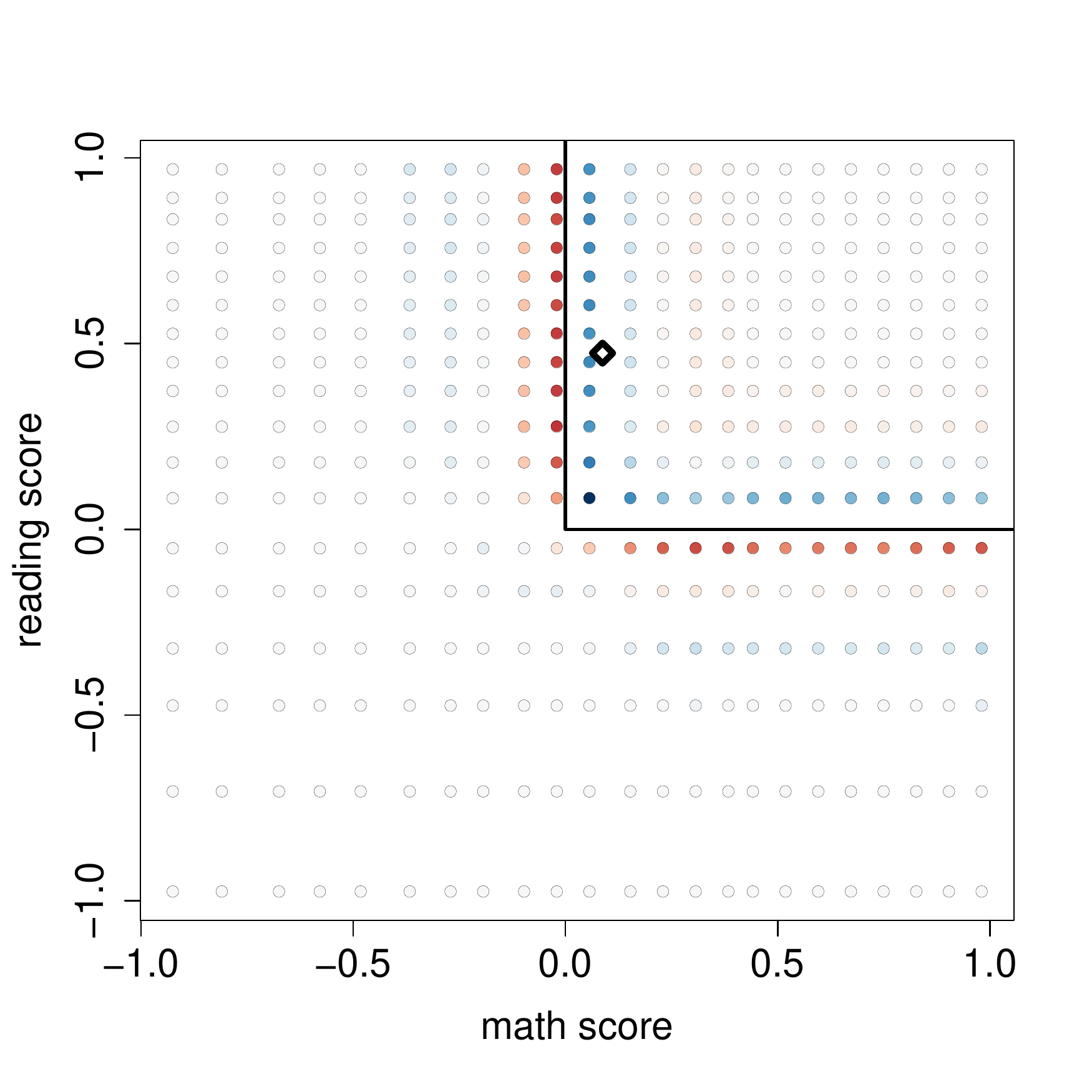} \\
unweighted CATE \eqref{eq:estimator_2d_pt} & weighted CATE \eqref{eq:estimator_2d_greedy}
\end{tabular}
\caption{Weights \smash{$\hgamma$} underlying treatment effect estimates of the effect of summer
school on the following year's reading scores, using both \eqref{eq:estimator_2d_pt} which seeks to
estimate the conditional average treatment effect (CATE) at $c = (0, \, 0)$, and the estimator
\eqref{eq:estimator_2d_greedy} which allows weighted CATE estimation. The size of $\hgamma_i$
is depicted by the color, ranging from dark red (very positive) to dark blue (very negative).
In the right panel, the diamond marks the weighted mean of the treated $X_i$-values, i.e., $\sum \hgamma_{i} W_i X_i$.
These plots were generated with a maximum second derivative bound of $B = 0.5 \times 40^{-2}$.}
\label{fig:2d_weights}
\end{figure}

We consider both of our optimized estimators, \eqref{eq:estimator_2d_pt} and \eqref{eq:estimator_2d_greedy},
and compare weight functions $\hgamma$ learned by both methods in Figure \ref{fig:2d_weights}.
The estimator $\htau_c$ is in fact quite conservative, and only gives large weights to students who
scored close to $c$. Our choice of estimating the conditional average treatment effect at
$(0, \, 0)$ may have been particularly challenging, as it is in
a corner of control-space and so does not have particularly many control neighbors.

In contrast, the weighted method $\htau_*$ appears to have effectively learned matching: It
constructs pairs of observations all along the treatment discontinuity, thus allowing it to
use more data while canceling out curvature effects due to $\mu_{0}(x)$.
As seen in Table \ref{tab:matsudaira}, in this sample, it is much more common to fail math and
pass reading than vice-versa; thus, the mean of the samples used for ``matching'' lies closer
to the math pass/fail-boundary than the reading one.

\setlength{\tabcolsep}{3pt}
\begin{table}[t]
\centering
\begin{tabular}{|r|c|ccc|ccc|}
\hline
\multicolumn{2}{|r|}{estimator: }
& \multicolumn{3}{c|}{unweighted CATE \eqref{eq:estimator_2d_pt}}
& \multicolumn{3}{c|}{weighted CATE \eqref{eq:estimator_2d_greedy}} \\ \hline
subject & $B$ & conf.~int. & m.~bias & samp.~err & conf.~int. & m.~bias & s.~err \\
\hline
math & $0.5 \times 40^{-2}$ & $0.037 \pm 0.093$ & 0.030 & 0.038 & $0.076 \pm 0.037$ & 0.009 & 0.017 \\
math & $1.0 \times 40^{-2}$ & $0.013 \pm 0.126$ & 0.041 & 0.052 & $0.068 \pm 0.043$ & 0.011 & 0.019 \\
\hline
reading & $0.5 \times 40^{-2}$ & $0.014 \pm 0.098$ & 0.030 & 0.041 & $0.044 \pm 0.037$ & 0.009 & 0.017 \\
reading & $1.0 \times 40^{-2}$ & $-0.015 \pm 0.130$ & 0.040 & 0.054 & $0.047 \pm 0.043$ & 0.011 & 0.019 \\
\hline
\end{tabular}
\caption{Estimates for the effect of summer school on math and reading scores on the following year's test,
using different estimators and choices of $B$. Reported are bias-adjusted 95\% confidence intervals, a bound
on the maximum bias given our choice of $B$, and an estimate of the sampling error conditional on $\cb{X_i}$.}
\label{tab:2d_results}
\end{table}
\setlength{\tabcolsep}{6pt}

In order to proceed with our approach, we again need to choose a value for $B$.
Running a 2nd order polynomial regression on the next year's math and reading scores for both treated and control
students separately, we find the largest curvature effect among the reading score of control students; roughly
a curvature of $0.46 \times 40^{-2}$ along the $(1, 2)$ direction. Thus, we run our algorithm with both an
optimistic choice of $B = 0.5 \times 40^{-2}$ and a more conservative choice $B= 1.0 \times 40^{-2}$
(we report curvatures on the ``scale'' of the plots in Figure \ref{fig:2d_weights}, such that a curvature of
$1.0 \times 40^{-2}$ results in a worst-case bias of $1$ in the corners of the plot).

Results are given in Table \ref{tab:2d_results}. As expected, the confidence intervals using the weighted
method \eqref{eq:estimator_2d_greedy} are much shorter than those obtained using \eqref{eq:estimator_2d_pt},
allowing for a 0.95-level significant detection in the first case but not in the second. Since the weighting
method allows for shorter confidence intervals, and in practice seems to yield a matching-like estimator,
we expect it to be more often applicable than the unweighted estimator \eqref{eq:estimator_2d_pt}. 

\begin{figure}[t]
\centering
\begin{tabular}{ccc}
\includegraphics[width=0.3\textwidth, trim = 10mm 0mm 10mm 5mm]{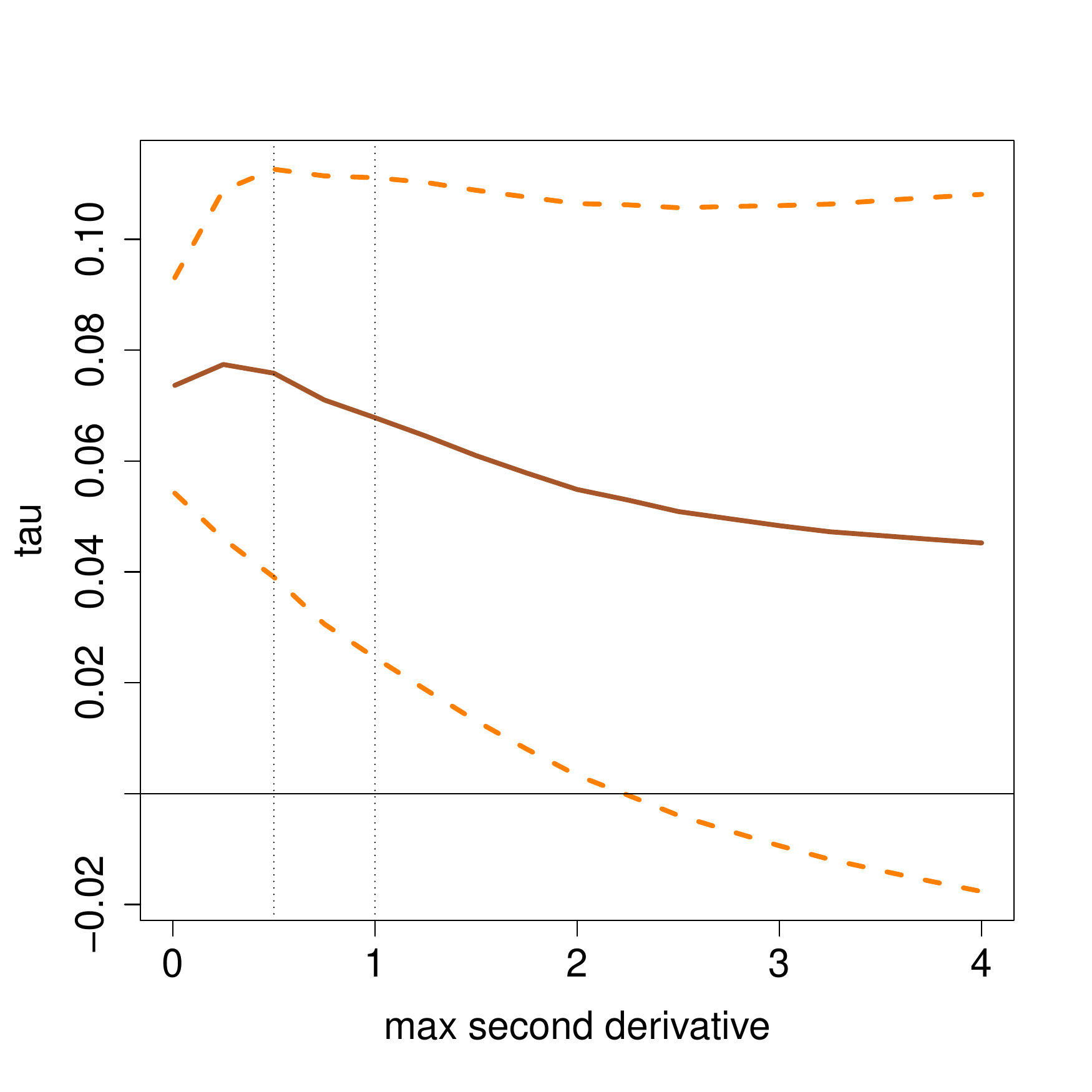} &
\includegraphics[width=0.3\textwidth, trim = 10mm 0mm 10mm 5mm]{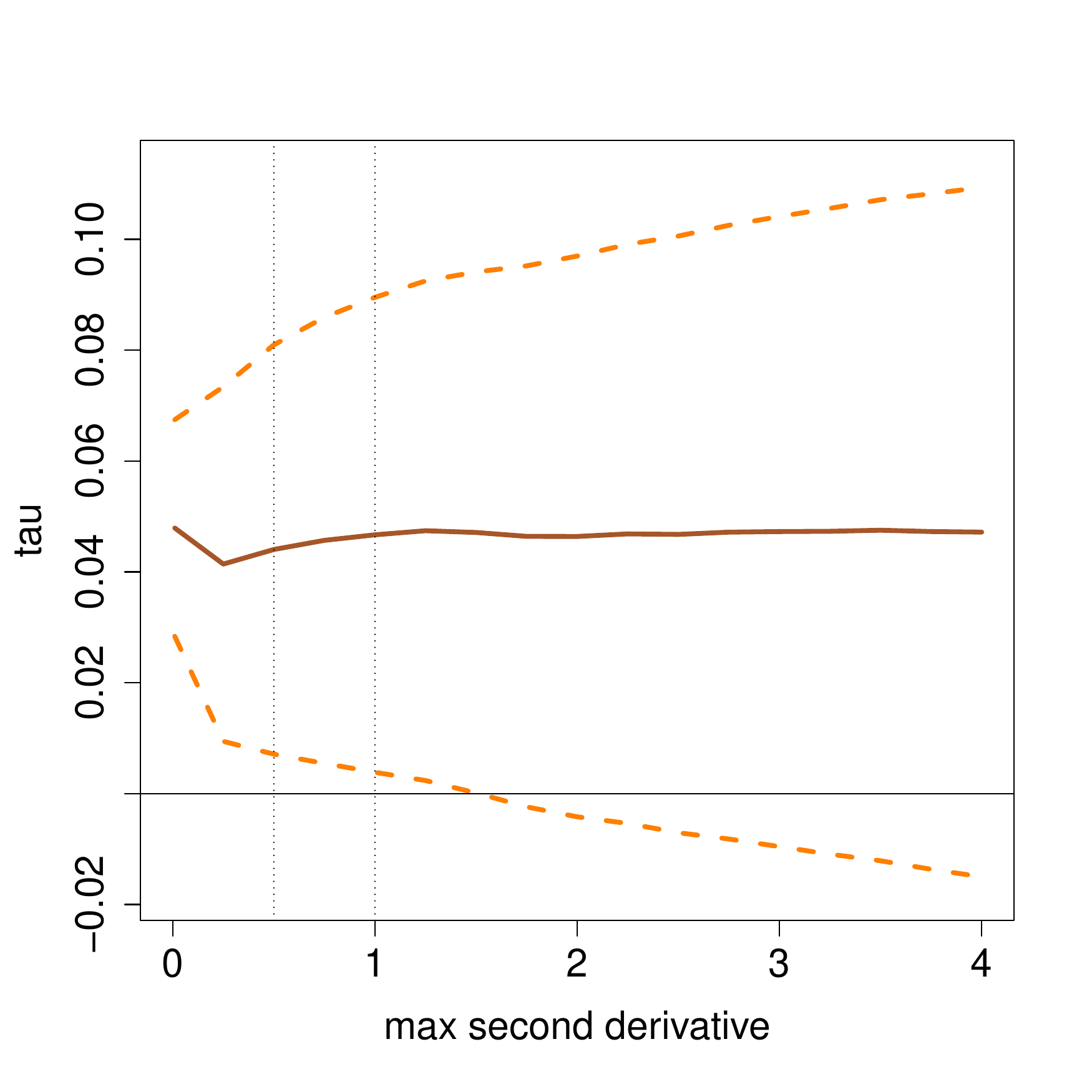} &
\includegraphics[width=0.3\textwidth, trim = 10mm 0mm 10mm 5mm]{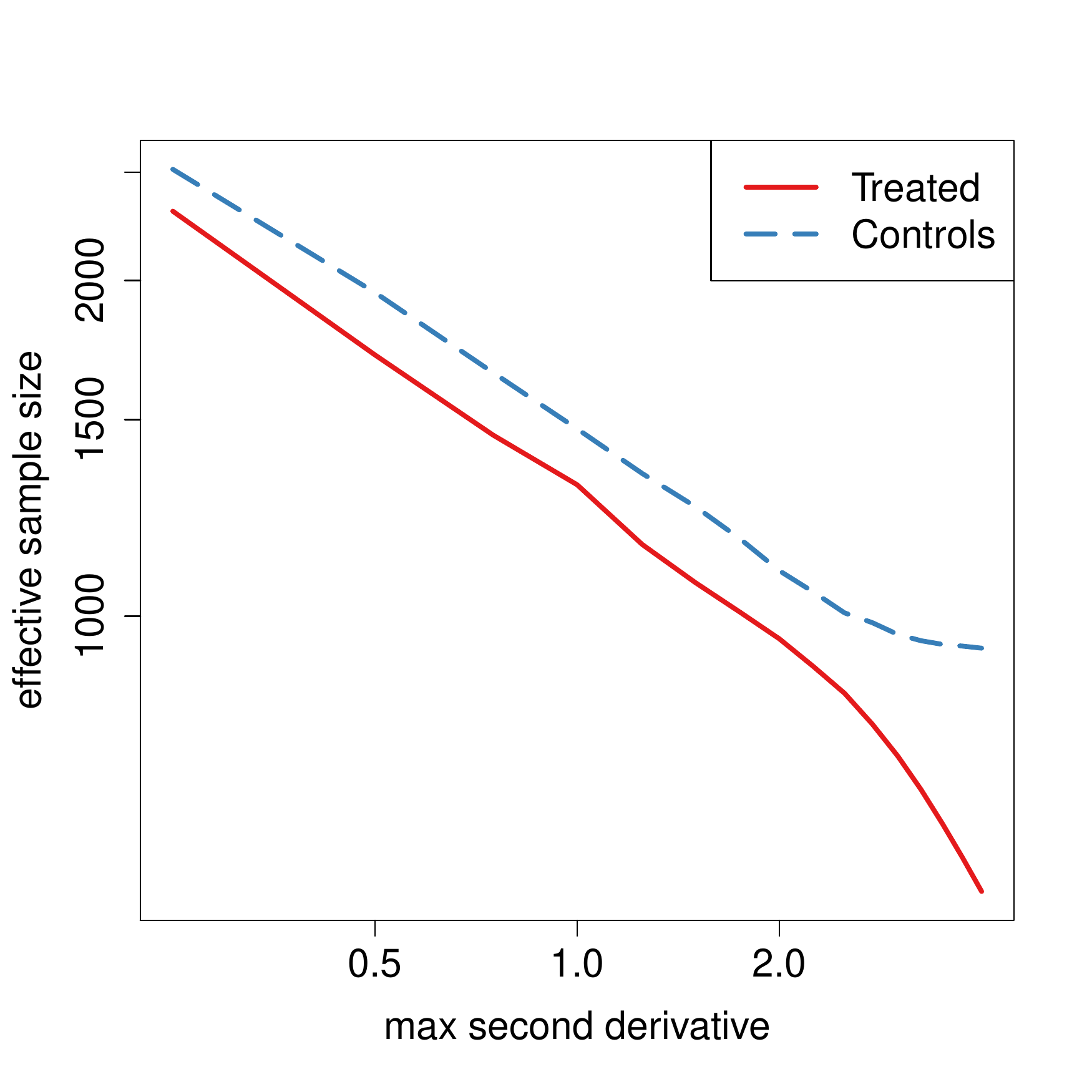} \\
Math Conf.~Interval & Reading Conf.~Interval & Effective Samp.~Size
\end{tabular}
\caption{The first two panels depict a sensitivity analysis for our weighted CATE result, for the math and reading
outcomes respectively. We plot point estimates along with bias-aware 95\% confidence intervals for different
choices of $B$; the choices of $B$ used in Table \ref{tab:2d_results} are indicated with dotted lines. The third
panel depicts effective sample sizes used by the procedure, \smash{$\text{ESS}_w = 1 / \sum_{\cb{i : W_i = w}} \hgamma_i^2$}.
For a given value of $B$, the \smash{$\hgamma_i$} used for the math and reading outcomes
are the same. In all cases, $B$ is multiplied by \smash{$40^{2}$} for readability.}
\label{fig:supplem_fig}
\end{figure}

Figure \ref{fig:supplem_fig} presents some further diagnostics for our result. The first two panels depict
a sensitivity analysis for our weighted CATE result:
We vary the maximum bound $B$ on the second derivative,
and examine how our confidence intervals change.\footnote{We note that, if the CATE function $\tau(\cdot)$
is not constant, then our weighted CATE estimand \smash{$\bar{\tau}_* = \sum_i W_i \hgamma_i \tau(X_i)$}
may vary with $B$. The result in Figure \ref{fig:supplem_fig} should thus formally be interpreted as either
a sensitivity analysis for the constant treatment effect parameter $\tau$ if we are willing to assume
constant treatment effects, or as a robustness check for our rejection of the global null hypothesis
$\tau(x) = 0$ for all $x$.}
The result on the effect of summer school on math scores
appears to be fairly robust, as we still get significant bias-aware 95\% confidence intervals at 
$B = 2 \times 40^{-2}$, which is 4 times the largest apparent curvature observed in the data. The third panel
plots a measure of the effective size of the treated and control samples used by the algorithm,
\smash{$\text{ESS}_w = 1 / \sum_{\cb{i : W_i = w}} \hgamma_i^2$}. Although our analysis sample has almost
exactly the same number of treated and control units (\smash{$\bW = 0.501$}), it appears that our algorithm
is able to make use of more control than treated samples, perhaps because the treated units are ``wrapped around''
the controls.

Finally, it is of course natural to ask whether the bivariate specification considered here gave us
anything in addition to the simpler approach used by \citet{matsudaira2008mandatory}, i.e., of estimating
the treatment effect of summer school on the next year's math exam by running a univariate regression
discontinuity analysis on only those students who passed the reading exam, and vice-versa to the effect on
the reading exam.\footnote{The estimator of \citet{matsudaira2008mandatory} is not exactly
comparable to the two we consider. For example, when only focusing on students who passed the
reading exam, his estimator effectively averages treatment effects over the math pass/fail boundary but not
the reading pass/fail boundary. In contrast, we either estimate the conditional average treatment effect
at a point $c$ \eqref{eq:estimator_2d_pt}, or allow for averaging over the full boundary
\eqref{eq:estimator_2d_greedy}. It is unclear whether the restriction of
\citet{matsudaira2008mandatory} to averaging over only one of the two boundary segments targets
a meaningfully more interpretable estimand than \eqref{eq:estimator_2d_greedy}.}
We ran both of these analyses using our method \eqref{eq:qp_general}, again considering bounds
$B = 0.5 \times 40^{-2}, \ 1 \times 40^{-2}$ on the second derivative.
For math, we obtained 95\% confidence intervals of
$0.083 \pm 0.040$ and $0.079 \pm 0.047$
for the smaller and larger $B$-bounds respectively; for reading, we obtained
$0.037 \pm 0.075$ and $0.030 \pm 0.090$.
In both cases, the corresponding bounds for the weighted estimator \eqref{eq:estimator_2d_greedy}
in Table \ref{tab:2d_results} are shorter, despite accounting for the possibility of bivariate curvature
effects.
The difference is particularly strong for the reading outcome, since our estimator $\htau_*$ can also use
students near the math pass/fail-boundary for improved precision.\footnote{The corresponding headline
numbers from \citet{matsudaira2008mandatory} are a 95\% confidence interval of $0.093 \pm 0.029$ 
for the effect on the math score, and $0.046 \pm 0.045$ for the reading score; see Tables 2 and 3,
reduced form estimates for 5th graders. These confidence intervals, however, do not formally account for
bias. They estimate the discontinuity parameter using a global cubic fit; such methods, however, do
not reliably eliminate bias \citep{gelman2014high}.}

\subsection{A Geographic Discontinuity Design}
\label{sec:keele_titiunik}

Our final application expands on the example of \citet{keele2014geographic}, briefly discussed
in Section \ref{sec:intro_multi}, the goal of which is to estimate the effect of political advertising
on participation in presidential elections by comparing households straddling a media market boundary.
As discussed in \citet{keele2014geographic}, inference hinges on the fact that they found a media
market boundary that appears not to coincide with any other major administrative boundaries, thus making
it more plausible that discontinuous responses across the geographic boundary are in fact caused by varying
media exposure.

The way \citet{keele2014geographic} approach this problem is that, given a focal point $c$, they first
measure Euclidean distance of each sample to it, $D_i = \Norm{X_i - c}_2$, and then use $D_i$ as a running
variable in a univariate regression discontinuity analysis. They establish consistency of their approach via a
local regression argument following \citet{hahn2001identification}, and also discuss advantages of their approach
relative to one that considers shortest distance to the treatment boundary (rather than $c$) as input
to a univariate analysis. Beyond consistency arguments, however, the resulting estimator has no
guarantees in terms of statistical optimality.
For example, the approach of \citet{keele2014geographic} collapses all treated (and respectively control)
observations at distance $D_i$ from $c$ to the same point in the univariate representation, which may increase bias.

\begin{figure}[t]
\centering
\begin{tabular}{cc}
\includegraphics[width=0.45\textwidth, trim=7mm 7mm 7mm 7mm]{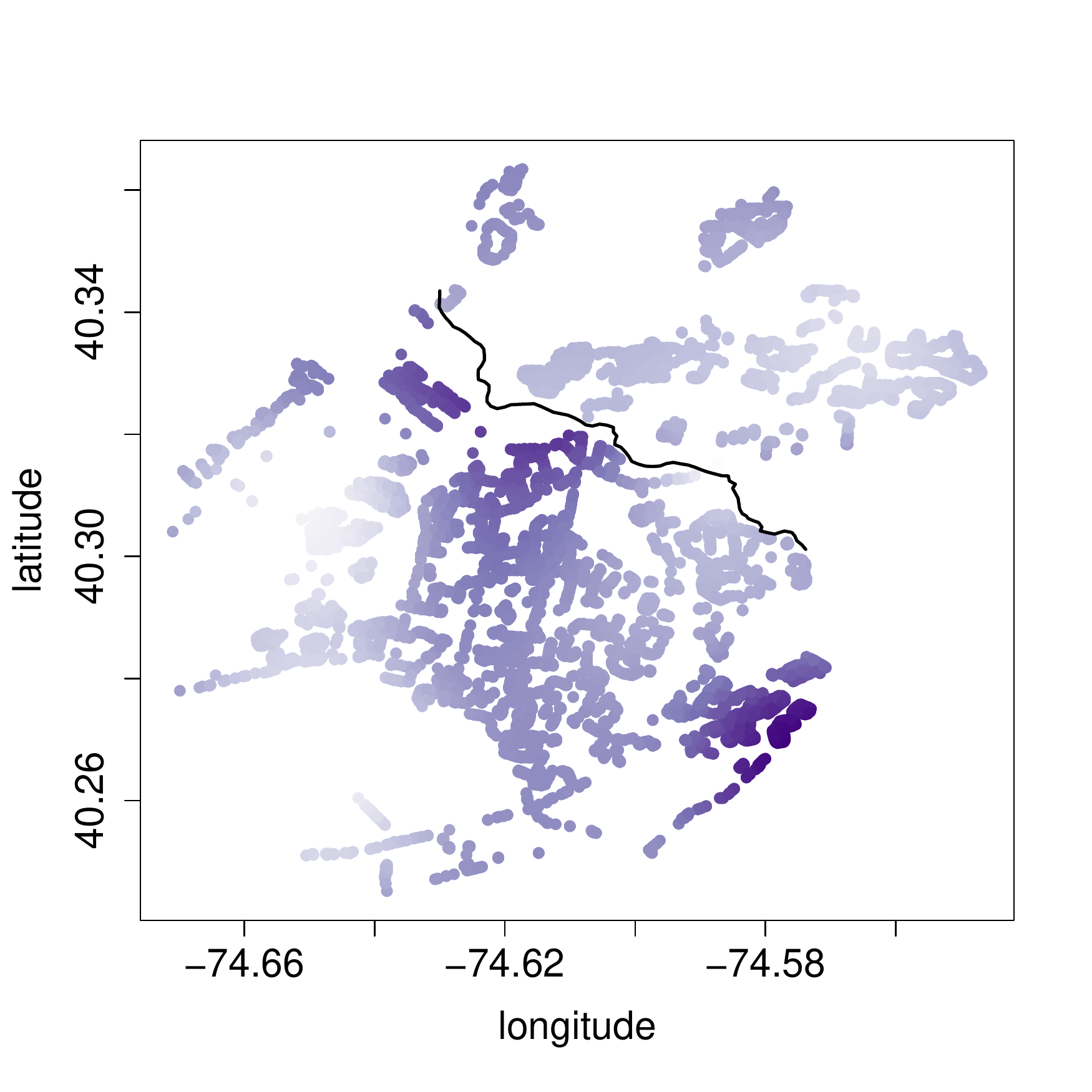} &
\includegraphics[width=0.45\textwidth, trim=7mm 7mm 7mm 7mm]{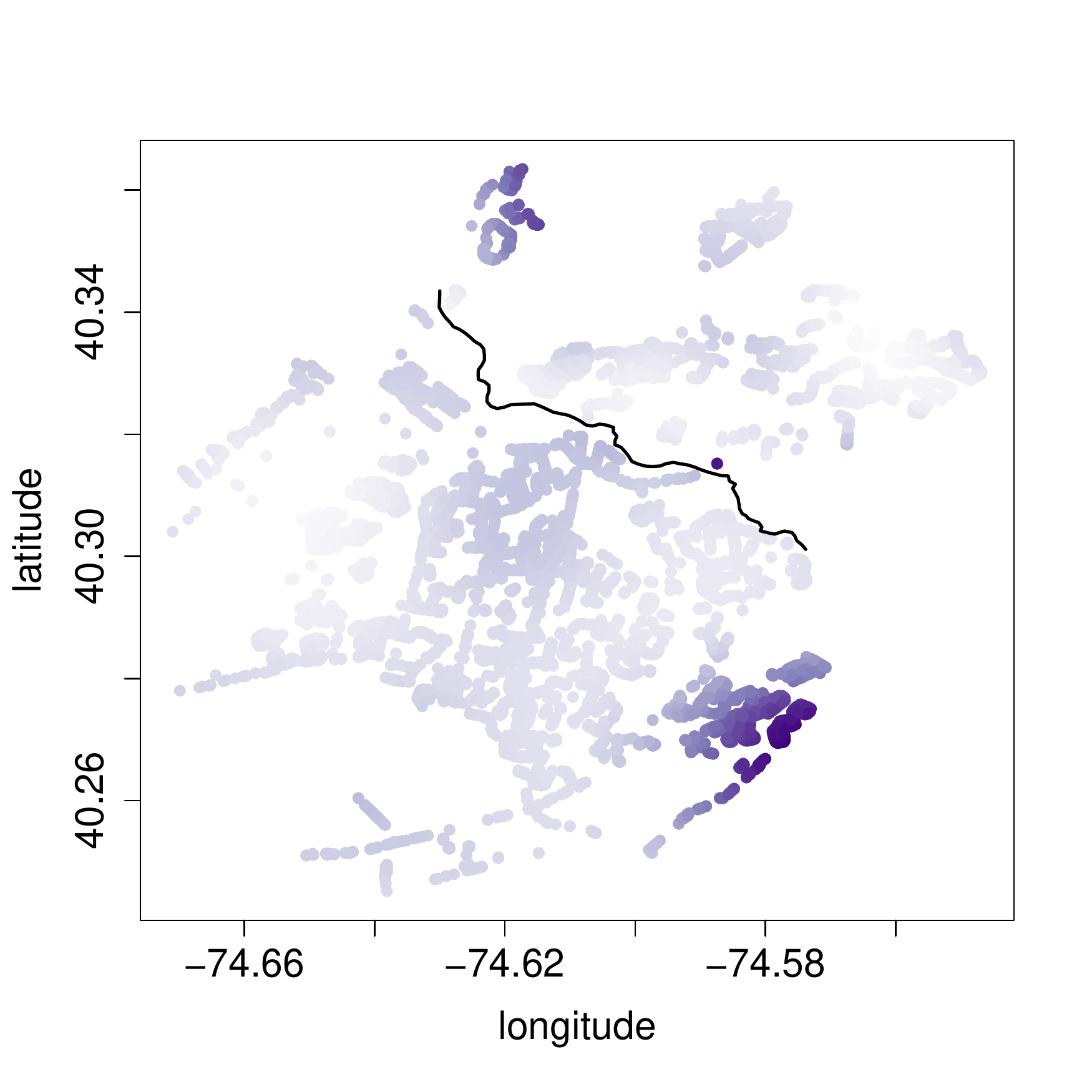}
\end{tabular}
\caption{Non-parametric estimates for turnout in the 2008 presidential elections (left panel)
and average age (right panel), as a function of geographic covariates.
There is considerable variation in the underlying signal:
localized turnout rates range from 26\% to 46\%,
while localized age averages range from 43 years to 68 years.
Darker regions correspond to larger values of the response.
These predictive surfaces were used to pick the curvature bound $B$.}
\label{fig:geo_curv}
\end{figure}

We apply our method to this problem and, following \citet{keele2014geographic}, seek to measure the effect of media exposure
on our main outcome of interest (participation in the 2008 presidential elections), as well as
placebo outcomes (age, black race, political party affiliation, and gender).
There were $n = 24,460$ observations in this school district.
Relative to the other examples considered in our paper, the responses here have a much richer
dependence on the geographic running variables, thus making the problem of bounding $B$ much more
delicate. To illustrate this issue, Figure \ref{fig:geo_curv} depicts non-parametric estimates of expected
turnout and age as a function of spatial location, and reveals strong local structure. For
example, we see two regions in the south-eastern and northern edges of the school district with a notably
higher average age than elsewhere.

We did not find simple quadratic methods for bounding $B$ to be sufficiently sensitive to this strong local structure,
and thus chose instead to set $B$ as the worst-case local curvature of a non-parametric global fit.
Specifically, we ran a cross-validated ridge regression with interacted 7th-order natural splines as features
on each side of the media market boundary, and then set $B$ at the 95-th percentile of the estimated
operator norm curvature over all the training points. Cross-validation was carried out via the \texttt{1.se} rule in
the \texttt{R}-package \texttt{glmnet} \citep{friedman2010regularization}.
We caution that this type of local choice of $B$ is only possible on fairly large datasets with strong local effects.
With smaller datasets or weaker signals, cross-validation may over-smooth the estimated response function,
thus resulting in an over-optimistically small value for $B$.

\setlength{\tabcolsep}{3pt}
\begin{table}[t]
\centering
\begin{tabular}{|r|r|cccc|}
  \hline
 & outcome & pt.~estimate & conf.~interval & max.~bias & std.~err. \\ 
  \hline
  \parbox[t]{2.8mm}{\multirow{5}{*}{\rotatebox[origin=c]{90}{$\tau$ at point}}}
 & 2008 turnout & 0.130 & (-0.072, 0.331) & 0.071 & 0.079 \\ 
   & age & 8.605 & (-8.949, 26.158) & 6.552 & 6.682 \\ 
   & black & -0.006 & (-0.080, 0.069) & 0.032 & 0.026 \\ 
   & democrat & 0.019 & (-0.080, 0.118) & 0.035 & 0.039 \\ 
   & female & 0.065 & (-0.060, 0.191) & 0.045 & 0.049 \\ 
   \hline
   \parbox[t]{2.8mm}{\multirow{5}{*}{\rotatebox[origin=c]{90}{weighted $\tau$}}}
   & 2008 turnout  & 0.045 & (-0.107, 0.197) & 0.052 & 0.060 \\ 
   & age & 3.105 & (-8.702, 14.913) & 4.147 & 4.649 \\ 
   & black & 0.016 & (-0.043, 0.075) & 0.024 & 0.021 \\ 
   & democrat & -0.027 & (-0.110, 0.055) & 0.026 & 0.034 \\ 
   & female & 0.019 & (-0.083, 0.120) & 0.033 & 0.042 \\ 
   \hline
\end{tabular}
\caption{Estimates for the effect of television advertising on voter turnout, as well as
placebo outcomes (age in years, black indicator, Democrat indicator, and female indicator).
The upper half of the table displays estimates for the conditional average treatment effect at
the point marked X in Figure \ref{fig:geo_weights}, whereas the lower half allows for
a weighted treatment effect as in \eqref{eq:estimator_2d_greedy}. We provide a point estimate for the treatment effect,
a 95\% confidence interval, as well as the worst case bias and sampling error used to produce the
confidence interval.}
\label{tab:geo_results}
\end{table}
\setlength{\tabcolsep}{6pt}

Table \ref{tab:geo_results} shows results from our analysis, both in terms of pointwise estimates
following \eqref{eq:estimator_2d_pt} and weighted estimates \eqref{eq:estimator_2d_greedy}.
We set the curvature parameter $B$ separately for each different outcome, following the procedure
outlined above.\footnote{\citet{keele2014geographic} also had a placebo outcome indicating Hispanic
ethnicity. However, this outcome does not appear to have much local structure,
and cross-validated ridge regression learned a constant function (resulting in $B = 0$). Thus, we
omit this placebo check from our present analysis; another alternative would have been to fall
back on our previous approach and to select $B$ via a global quadratic fit.}
The $\gamma$-weighting functions underlying our analysis for the primary outcome are shown
in Figure \ref{fig:geo_weights}.
Overall, we replicate the finding of \citet{keele2014geographic}, in
that the media market boundary does not appear to have a substantial association with either the
primary outcome of the placebo outcomes.
Relative to existing methods, the value of our approach is that it allows for transparent inference
of causal parameters using the original untransformed data (rather than relying on a hand-constructed univariate
running variable), and uniformly accounts for spatial curvature effects.

\section{Discussion}
\label{sec:discussion}

In this paper, we introduced an optimization-based approach to statistical inference in regression
discontinuity designs. By using numerical convex optimization tools, we explicitly derive the
minimax linear estimator for the regression discontinuity parameter under bounds on the
second derivative of the conditional response surface. Because any method based on local
linear regression is also a linear estimator of this type, our approach dominates local linear
regression in terms of minimax mean-squared error. We also show how our approach
can be used to build uniformly valid confidence intervals.

A key advantage of our procedure is that, given bounds on the second derivative, estimation
of the regression discontinuity parameter is fully automatic. The proposed algorithm is the
same whether the running variable is continuous or discrete, and whether or not $\tau(c)$ is
point identified. Moreover, it does not depend on the
shape of treatment assignment boundary when $X$ is multivariate. We end our discussion with
some potential extensions of our approach.

\paragraph{Fuzzy regression discontinuities}

In the present paper, we only considered sharp regression discontinuities, where the treatment
assignment $W_i$ is a deterministic function of $X_i$. However, there is also considerable interest
in fuzzy discontinuities, where $W_i$ is random but $\PP{W_i = 1 \cond X_i = x}$ has a jump at
the threshold $c$; see \citet{imbens2008regression} for a review. In this case, it is common
to interpret the indicator $\ind\p{\cb{X_i \geq c}}$ as an instrument, and then to estimate
a local average treatment effect via two-stage local linear regression \citep{imbens1994identification}.
By analogy, we can estimate treatment effects with fuzzy regression discontinuity
via two-stage optimized designs as
\begin{equation}
\htau_{LATE} = \sum_{i = 1}^n \hgamma_i Y_i \ \Big/\ \sum_{i = 1}^n \hgamma_i W_i,
\end{equation}
where the $\hgamma_i$ are obtained as in \eqref{eq:qp_general} with an appropriate choice penalty on
the maximal squared imbalance $t^2$. This approach would clearly be consistent based on results
established in this paper; however, deriving the best way to trade off bias and variance in
specifying $\hgamma_i$ and extending the approach of \citet{donoho1994statistical} for
uniform asymptotic inference is left for future work.

\paragraph{Balancing auxiliary covariates}

In many applications, we have access to auxiliary covariates $Z_i \in \RR^p$ that are
predictive of $Y_i$ but unrelated to the treatment assignment near the boundary $c$.
As discussed in, e.g., \citet{imbens2008regression}, such covariates are not necessary
for identification; but controlling for them can increase robustness to hidden biases.
One natural way to use such auxiliary covariates in our optimized designs is to require
the weights \smash{$\hgamma_i$} to balance these covariates, i.e., to add a constraint
\begin{equation}
\label{eq:balance}
\sum_{i = 1}^n \hgamma_i \, Z_{ij} = 0 \text{ for all } j = 1, \, ..., \, p
\end{equation}
to the optimization problem \eqref{eq:estimator}.\footnote{The constraint \eqref{eq:balance}
is a linear constraint, and so the optimization problem \eqref{eq:qp_general} remains a quadratic
program with this constraint.} In principle, if the distribution of $Z_i$ is in fact independent
of $X_i$ when $X_i$ is near the threshold $c$, we would expect the balance conditions \eqref{eq:balance}
to hold approximately even if we do not enforce them;
however, explicitly enforcing such balance may improve robustness.\footnote{A related idea
would be to use the covariates $Z_i$ for post-hoc specification testing as in
\citet{heckman1989alternative} or \citet{imbens2008regression}. Their strategy is to obtain
weights $\hgamma_i$ without looking at the $Z_i$, and then to reject the modeling
strategy if \eqref{eq:balance} does not hold approximately.}
If we have an additive, linear dependence of $Y_i$ on $Z_i$, then enforcing balance
as in \eqref{eq:balance} would also result in variance reduction, as the conditional
variance of our estimator \smash{$\htau$} would now depend on
$\Var{Y_i \cond X_i, \, Z_i}$, which is always smaller or equal to $\Var{Y_i \cond X_i}$.

\paragraph{Working with generic regularity assumptions}

Following standard practice in the regression discontinuity literature, we
focused on minimax linear inference under bounds on the second derivative
of $\mu_w(\cdot)$ \citep[e.g.,][]{kolesar2016inference,imbens2011optimal}.
However, our conceptual framework can also be applied with higher order
smoothness assumptions via bounds on the $k$-th derivative of $\mu_w(\cdot)$,
and can easily be combined with other forms of structural information about
the conditional response functions (e.g., perhaps we know from theory that the functions
$\mu_w(\cdot)$ must be concave). Thanks to the flexibility of our optimization-based
approach, acting on either of these ideas would simply involve implementing
the required software using standard convex optimization libraries.

\bibliographystyle{plainnat}
\bibliography{references}

\section{Appendix: Proofs}

\subsection{Proof of Theorem \ref{theo:gauss}}

By construction, we already know that
$$ \EE{\htau \cond X_1, \, ..., \, X_n}  - \tau(c) = b\p{\hgamma}, \ \ \Var{\htau \cond X_1, \, ..., \, X_n} = s^2\p{\hgamma}. $$
Thus, to establish our desired result, it suffices to establish asymptotic Gaussianity of
$$ \htau - \EE{\htau \cond X_1, \, ..., \, X_n} = \sum_{i = 1}^n \hgamma_i \varepsilon_i, \ \ \varepsilon_i = Y_i - \mu_{W_i}(X_i). $$
To do so, we use the Lyapunov central limit theorem. Thanks to our bound on the $q$-th moment
of the $\varepsilon_i$, we know that
\begin{align*}
&\sum_{i = 1}^n \EE{\p{\hgamma_i \varepsilon_i}^q \cond \hgamma_i} \, \Big/ \, s^q\p{\hgamma}
\leq \sum_{i = 1}^n C\hgamma_i^q \, \Big/ \, \p{\sigma_{\min}^q \Norm{\hgamma}_2^q} \\
&\ \ \ \ \ \ \leq \frac{C}{\sigma_{\min}^q} \sup_{1 \leq i \leq n}\cb{\abs{\hgamma_i} \big/ \Norm{\hgamma}_2}^{q - 2}
\rightarrow_p 0
\end{align*}
by assumption \eqref{eq:gamma_sup}. Thus, in particular, there exists a sequence $a_n$ such that $a_n \rightarrow 0$ and
$$ \limsup_{n \rightarrow \infty} \ \PP{\sum_{i = 1}^n \EE{\p{\hgamma_i \varepsilon_i}^q \cond \hgamma_i} \, / \, s^q\p{\hgamma} > a_n} = 0. $$
Now, define weights $\tgamma_i$ such that $\tgamma = \hgamma$ whenever
\smash{$\sum_{i = 1}^n \mathbb{E}[(\hgamma_i \varepsilon_i)^q \cond \hgamma_i] \, \Big/ \, s^q(\hgamma) \leq a_n$},
and $\tgamma_i = 1$ for all $i = 1, \, ..., \, n$ else. By Lyapunov's central limit theorem,
$\sum_{i =1}^n \tgamma_i \varepsilon_i \,/\, (\sum_{i = 1}^n \gamma_i^2 \sigma_i^2)^{1/2}$
is asymptotically standard normal. Moreover, we know that $\tgamma = \hgamma$ with
probability tending to 1, and so our estimator $\htau$ must also be conditionally standard normal
as claimed.

\subsection{Proof of Proposition \ref{prop:approx}}

Our proposed approximation to \eqref{eq:dual_simplified} has two main components. First, we need
to handle the second derivative constraints $\lVert\nabla^2 \tf_w(x)\rVert \leq \lambda_1$ for all values
of $x \in \xx$. One approach would be to write this as a positive semidefinite matrix constraint, i.e.,
$ \nabla^2 \tf_w(x) \preceq \lambda_1 \ii_{k \times k}$, and then rely on semidefinite programming tools;
here, however, we find that we can effectively approximate $\lVert\nabla^2 \tf_w(x)\rVert \leq \lambda_1$
via a finite number of linear constraints. Second, we need address the effect of finite differencing on our
optimization.

For simplicity, throughout the following discussion, we take the curvature bound $\lambda_1$ to be
fixed, and only consider minimization over $\lambda_2$, ..., $\lambda_5$, $G$, and the $\tf_w$.
Given this assumption, as well as the fact that $\xx$ is bounded, we can verify that our objective
is Lipschitz, which is helpful in bounding approximation errors. Then,
after establishing that our approximation scheme is valid for any fixed value of $\lambda_1$, it follows
that the approximation is also valid when we optimize over $\lambda_1$ because our objective is
strongly convex in $\lambda_1$.

We start by approximating the curvature constraint.
Let $A$ be any positive semidefinite matrix, and
let $\vv = \cb{v_1, \, ..., \, v_l}$ be a set of vectors.
Then, the spectral norm of $A$ is bounded by
\begin{equation}
\label{eq:spectral}
\begin{split}
&\Norm{A}_{\vv} \leq \Norm{A} \leq \p{1 - \alpha^2(\vv)}^{-1} \Norm{A}_{\vv}, \where \\
&\Norm{A}_{\vv} := \sup\cb{v^\top A v \,\big/\, \Norm{v}_2^2: v \in \vv} \eqand \\
&\alpha^2(\vv) = \sup\cb{\inf\cb{1 - \p{u \cdot v}^2 \,/\, \Norm{v} : v \in \vv} : \Norm{u}_2 = 1};
\end{split}
\end{equation}
note that $\alpha(\vv)$ measures the sine of the worst angular error we may make by approximating
a vector $u$ with an element from $\vv$. The first inequality in \eqref{eq:spectral} follows
from the definition of the spectral norm, while the second is immediate by expressing elements
in $\vv$ in the spectral basis of $A$.

In our finite dimensional approximation to \eqref{eq:dual_simplified}, we propose replacing the spectral
norm constraints $\Norm{\cdot}$ with constraints of the type $\Norm{\cdot}_\vv$ described above.
To motivate this strategy, write $\law\p{\Norm{\cdot}, \, \lambda_1}$ for the objective value of
\eqref{eq:dual_simplified} given a fixed value of $\lambda_1$, and let $\law\p{\Norm{\cdot}_{\vv}, \, \lambda_1}$
denote the related quantity with the approximating constraint.

Because the value of the objective cannot get worse as we increase the size the size of the constraint set,
\eqref{eq:spectral} implies that for any value of $\lambda_1$,
$$ \law\p{\Norm{\cdot}_\vv, \, \lambda_1} \leq \law\p{\Norm{\cdot}, \, \lambda_1} \leq \law\p{\Norm{\cdot}_{\vv}, \, \p{1 - \alpha^2(\vv)} \lambda_1}. $$
Moreover, any feasible solution to \eqref{eq:dual_simplified} with bound $\lambda_1$ on the constraint
can be turned into a feasible solution with bound $c\lambda_1$ by simply multiplying
$\tf(\cdot)$ and $\lambda_2, \, ..., \, \lambda_5$ by $c$. When we do so, the positive term
$\sum \sigma_i^{-2} G_i^2$ term scales quadratically in $c$, while the negative term
$\lambda_2 - \lambda_3$ scales linearly in $c$ (the second term must be negative, since
otherwise we could improve on the objective by setting everything to 0); in particular, this
implies that
$$ \law\p{\Norm{\cdot}_{\vv}, \, \p{1 - \alpha^2(\vv)} \lambda_1} \leq \p{1 - \alpha^2(\vv)} \law\p{\Norm{\cdot}_\vv, \, \lambda_1}. $$
Chaining these two inequalities together, we find that
$$  \law\p{\Norm{\cdot}_\vv, \, \lambda_1} \leq \law\p{\Norm{\cdot}, \, \lambda_1} \leq \p{1 - \alpha^2(\vv)} \law\p{\Norm{\cdot}_\vv, \, \lambda_1}, $$
and so, given a small enough value of $\alpha^2(\vv)$, the values of the two objectives match.
Finally, because the objective of \eqref{eq:dual_simplified} is strongly convex in
\smash{$\tf(X_1), \, ..., \, \tf(X_n)$},  $\lambda_2,\, \lambda_3,\, \lambda_4,\, \lambda_5$,
we find that if \smash{$ \law\p{\Norm{\cdot}_\vv, \, \lambda_1} - \law\p{\Norm{\cdot}, \, \lambda_1}$}
is goes to zero, then the solutions to the two optimization problems must also converge to each other.

\begin{figure}
\begin{center}
\begin{tabular}{cc}
\includegraphics[width=0.4\textwidth]{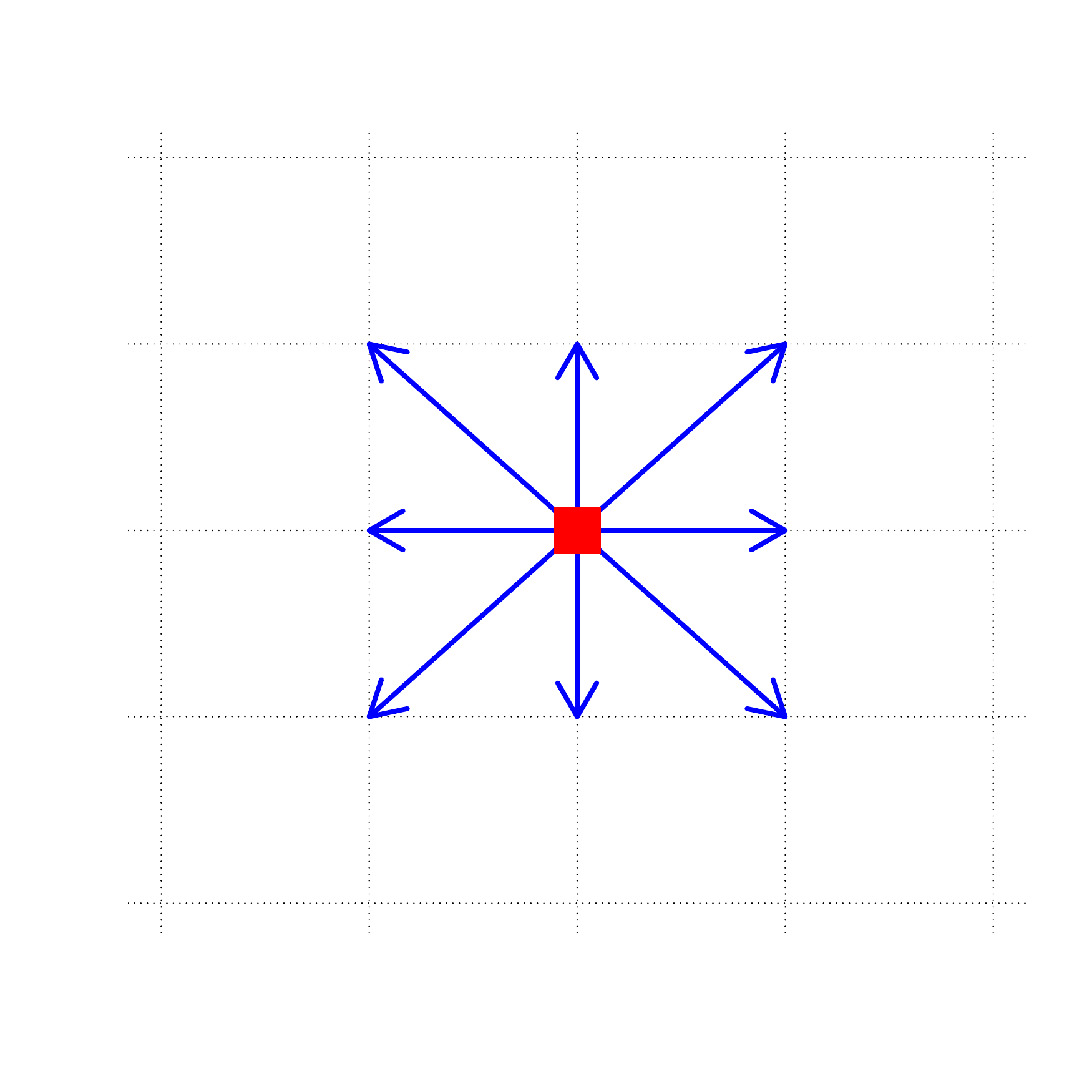}&
\includegraphics[width=0.4\textwidth]{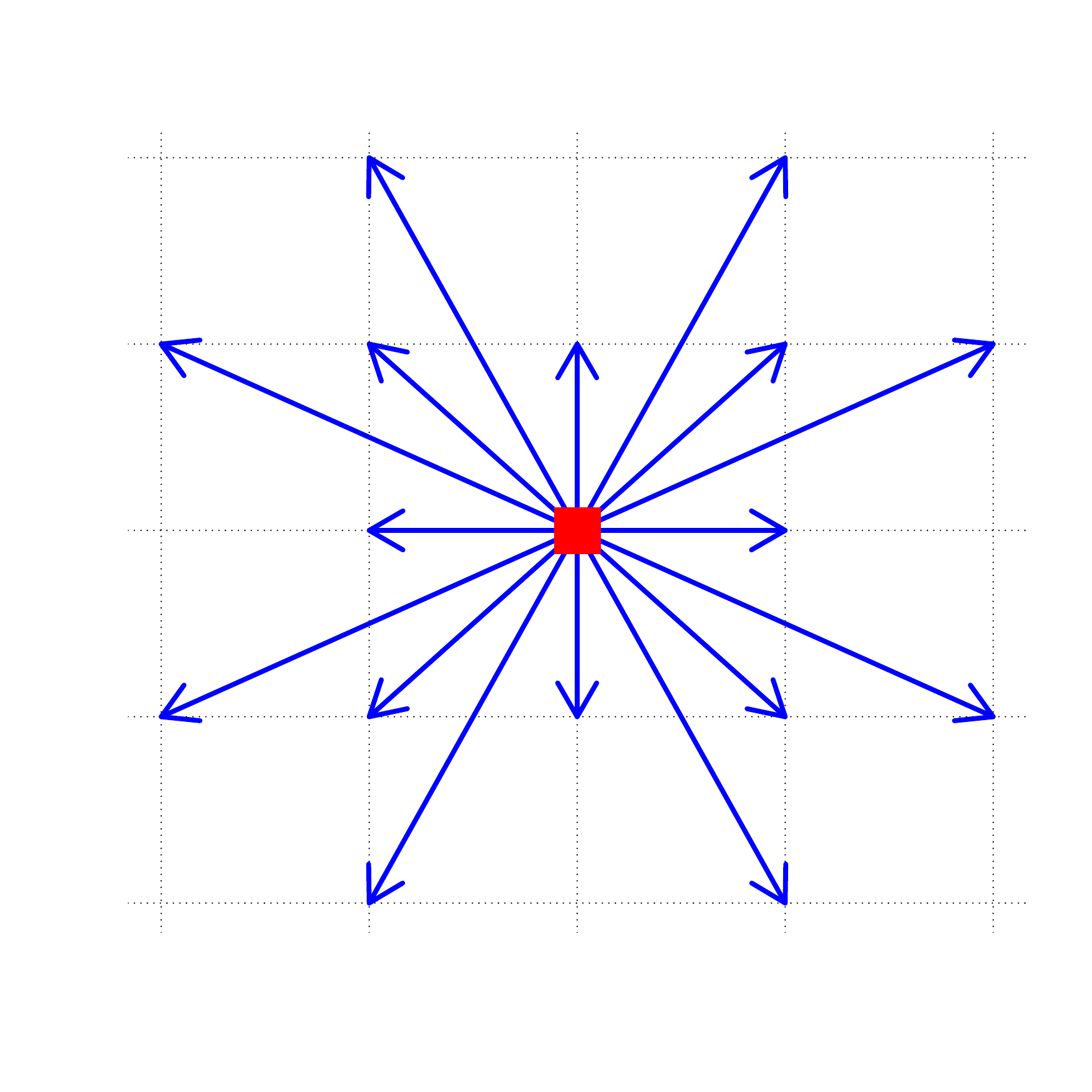}
\end{tabular}
\caption{Examples of grid-aligned approximating sets $\vv$ in $\RR^2$, with 4 and 8
(non-redundant) elements respectively.\label{fig:grid}}
\end{center}
\end{figure}

The remaining question is then to exhibit constructions of approximating sets $\vv$ that make $\alpha^2(\vv)$ small.
With a two-dimensional running variable, there exist grid-aligned sets $\vv$ with
4 and 8 elements respectively, depicted in Figure \ref{fig:grid}, that attain values of $\alpha^2(\vv)$ of
0.15 and 0.05 respectively. This construction naturally extends to arbitrary $k$, resulting in sets $\vv$
whose size scales polynomially in the inverse of the tolerance $\alpha^2$. In practice, though, this
other computational strategies may be needed when $k$ is larger than 2 or 3.

Our second task is in showing that we can solve the $\Norm{\cdot}_{\vv}$-norm
optimization problem using a finite difference approximation. Let $h$ be a bandwidth, and let
$\xx_h$ be the set of grid points of a lattice with edge-length $h$ that fall within $h$, and
assume that: the set $\vv$ consists only of vectors whose end points lie on the unit lattice (as in, e.g., Figure \ref{fig:grid}),
the basis vectors $e_j$ belong to $\vv$ for $j = 1, \, ..., \, k$,
and that the center point $c$ is in $\xx_h$.
Then, writing $N_h(x)$ for a mapping that takes all points in $\xx$ to one of the nearest points
in $\xx_h$, we consider the following finite-dimensional approximation
to \eqref{eq:dual_simplified}:
\begin{equation*}
\label{eq:finite_difference}
\begin{split}
&\text{minimize } \frac{1}{4} \sum_{i = 1}^n \sigma_i^{-2} G_i^2 + \frac{\lambda_1^2}{4B^2} + \lambda_2 - \lambda_3 \\
&\text{subject to } G_i = \tf^{(h)}(N_h(X_i)) + \lambda_2 w(X_i) + \lambda_3 (1 - w(X_i))  \\
&\ \ \ \ \ \ \ \ \ \ \ \ \ \ \ \ \ \ \ \ \ \ \ \ \ \ \ \ + \lambda_4 (X_i - c) + \psi \, \lambda_5 (2w(X_i) - 1) (X_i - c) \\
&\ \ \tf^{(h)}(x) = \tf_0^{(h)}(x) + \psi \, w(x) \p{\tf_1^{(h)}(x) - \tf_0^{(h)}(x)} \ \forall x \in \xx_h, \ \lambda_1 \geq 0, \ \lambda_2, \, \lambda_3 \in \RR, \ \lambda_4, \, \lambda_5 \in \RR^k, \\
&\ \ \tf_w^{(h)}(c) = 0, \ \tf_w^{(h)}(c + h e_j) = 0 \text{ for all } j = 1, \, ..., \, k, \, w \in \cb{0, \, 1}\\
&\ \ \Norm{hv}_2^{-2}\abs{\tf_w^{(h)}(x + hv) + \tf_w^{(h)}(x -  hv) - 2\tf_w^{(h)}(x)} \leq \lambda_1 \text{ for all } x \in \xx_h, \, v \in \vv, \, w \in \cb{0, \, 1}.
\end{split}
\end{equation*}
By analogy to our earlier notation, write $\law_h\p{\Norm{\cdot}_\vv, \, \lambda_1}$ for the objective
value in the above problem for a fixed value of $\lambda_1$.

We first bound  $\law_h\p{\Norm{\cdot}_\vv, \, \lambda_1}$ in terms of $\law\p{\Norm{\cdot}_\vv, \, \lambda_1}$.
To do so, simply note that any feasible solution to the continuous problem can be turned into a solution to the discrete
problem by setting
$$\tf^{(h)}_w(x) \leftarrow \tf_w(x) - h^{-1}\sum_{j = 1}^k \p{x_j - c_j}\tf\p{c + he_j}, $$
and making appropriate corrections to $\lambda_2, \, ..., \, \lambda_5$.
The goal of this affine correction is to ensure that the discrete derivate constraints
\smash{$\tf^{(h)}_w(c + h e_j) = 0$} are satisfied. Now, because $e_j \in \vv$ and
\smash{$\nabla \tf_w(c) = 0$}, we know that $\tf\p{c + he_j} \leq \lambda_1h^2/2$;
thus, because $\xx$ is compact, we find that
$\law_h\p{\Norm{\cdot}_\vv, \, \lambda_1} \leq \law\p{\Norm{\cdot}_\vv, \, \lambda_1} + \oo\p{h\lambda_1}$.

Now, to go the other direction, we start we a feasible solution \smash{$\tf^{(h)}_w(x)$} to the discrete
solution, and need to construct a feasible solution to the continuous problem without making the objective
much worse. To do so, let $\varphi_k(\cdot)$ for the standard Gaussian density in $\RR^k$, and define
$$ \tg_w(z) = \p{\frac{h}{\beta_h}}^k \sum_{x \in \xx_h} \varphi_k\p{\frac{x - z}{\beta_h}}\tf_w^{(h)}(x) + a_w \cdot z, $$
with $\beta_h = h^{1/4}$,
where $a_w$ is a linear correction term used to ensure that \smash{$\nabla \tg_w(c) = 0$}, and we again
adjust the parameters $\lambda_2, \, ..., \, \lambda_5$ appropriately.
Then we set \smash{$\tf_w(z) := \tg_w(z)$} for all points $z$ for which a ball of radius $h^{1/8}$
is contained inside $\xx$, and use Whitney's extension theorem to extend this function to the rest of
$\xx$ without increasing the supremum of the second derivative. In order to verify that
\smash{$\law\p{\Norm{\cdot}_\vv, \, \lambda_1} \leq \law_h\p{\Norm{\cdot}_\vv, \, \lambda_1} + o(1)$},
it suffices to show that the new \smash{$\tf_w(z)$} almost satisfy the desired curvature constraints with almost
as good an objective value as before; the rest of the argument then follows as usual.

As the a full discussion is rather technical, we here focus on simply showing how to bound the
second derivative of \smash{$\tg_w(z)$} along vectors $v \in \vv$ at a point $z$ in the interior of $\xx$.
We emphasize that the our construction of $\tg_w(z)$ and $\tf_w(z)$
from \smash{$\tf^{(h)}_w(x)$} is merely a ``proof of concept'' used to establish asymptotic
equivalence of two optimization problems, and that this construction plays no role in the actual algorithm
we use.

Given a point of interest $z$, it is helpful to partition the space $\xx_h$ into chords
indexed by $r \in \rr$, such that any point $x \in \xx_h$ can uniquely
be written as $x = a_r + \p{\delta_r + mh}v$ for some $r \in \rr$ and $m \in \ZZ$. Here, $a_r$ is the intersection of the chord
with the normal space of $v$ through $z$, and $|\delta_r| \leq h/2$ is the smallest correction term enabling such a representation.
Now, we see that
\begin{align*}
\frac{d^2}{dt^2} \tg_w(z + tv/\Norm{v}_2) =  \frac{h^k}{\beta_h^{k+2}}  \sum_{x \in \xx_h} \p{\frac{\langle x - z, \, v/\Norm{v}_2 \rangle^2}{\beta_h^2} - 1} \varphi_k\p{\frac{x - z}{\beta_h}}\tf_w^{(h)}(x)
\end{align*} 
and, decomposing this quantity into a sum over chords, we get that
\begin{align*}
&\frac{d^2}{dt^2} \tg_w(z + tv/\Norm{v}_2)  \\
&\ \ = \frac{h^k}{\beta_h^{k+2}} \sum_{r \in \rr} \sum_{m \in \set_r} \p{\frac{\p{\delta_r + mh}^2 \Norm{v}_2^2}{\beta_h^{2}} - 1}   \varphi_k\p{\frac{a_r + \p{\delta_r + mh} v}{\beta_h}} \\
&\ \ \ \ \ \ \ \ \ \ \cdot \tf_w^{(h)}(z + a_r + \p{\delta_r + mh} v),
\end{align*}
where $\set_r \subset \ZZ$ denotes the set of valid $m$ indices in the chord $r$.

Now, the reason we used a Gaussian kernel $\varphi$ is that it has independent orthogonal components, and 
so we can split the $k$-dimensional Gaussian density $\varphi_k$ into a $(k-1)$-dimensional
density over the space normal to $v$ and a $1$-dimensional density for the contribution along $v$:
\begin{align*}
&\cdots = \frac{h^k}{\beta_h^{k+2}} \sum_{r \in \rr} \varphi_{k-1}\p{\frac{a_r}{\beta_h}} \sum_{m \in \set_r} \p{\frac{\p{\delta_r + mh}^2 \Norm{v}_2^2}{\beta_h^{2}} - 1}   \varphi_1\p{\frac{\p{\delta_r + mh} \Norm{v}_2}{\beta_h}} \\
&\ \ \ \ \ \ \ \ \ \ \cdot \tf_w^{(h)}(z + a_r + \p{\delta_r + mh} v).
\end{align*} 
We are then ready to take a ``Taylor expansion'' where
the first summand below captures the contributions of a linear effect in each chord,
and the second chord has the resulting second-order terms:
\begin{align*}
&\cdots = \frac{h^k}{\beta_h^{k+2}} \sum_{r \in \rr} \varphi_{k-1}\p{\frac{a_r}{\beta_h}} \sum_{m \in \set_r} \p{\frac{\p{\delta_r + mh}^2 \Norm{v}_2^2}{\beta_h^2} - 1}   \varphi_1\p{\frac{\p{\delta_r + mh} \Norm{v}_2}{\beta_h}}\\
&\ \ \ \ \ \ \ \ \ \ \cdot \p{\tf_w^{(h)}(z + a_r + \delta_r v) + m\p{\tf_w^{(h)}(z + a_r + (\delta_r + h) v) - \tf_w^{(h)}(z + a_r + \delta_r v)}} \\
&\ \ \ \ \  + \frac{h^k}{\beta_h^{k+2}} \sum_{r \in \rr} \varphi_{k-1}\p{\frac{a_r}{\beta_h}} \sum_{m \in \set_r} \p{\frac{\p{\delta_r + mh}^2 \Norm{v}_2^2}{\beta_h^2} - 1}   \varphi_1\p{\frac{\p{\delta_r + mh} \Norm{v}_2}{\beta_h}} \Delta(r, \, m),
\end{align*} 
where the term \smash{$\Delta(r, \, m)$} simply denotes the residual from the 1-st order approximation.

If we didn't have any discretization, the $\sum_{m \in \set_r}$ sum
inside the term due to first-order effects would be exactly 0 for each $r$ individually. Here, the discretization causes
error on the order of at most $h$ at each sampling point while each chord has at most
on the order of $h^{-1}$ points, meaning that the total sum is bounded on the order of
$$ \frac{h^k}{\beta_h^{k+2}} \sum_{r \in \rr} \varphi_{k-1}\p{\frac{a_r}{\beta_h}} = \oo\p{\frac{h}{\beta_h^{3}}} = \oo\p{h^{1/4}} $$
given our chosen growth rate on $\beta_h$.
We also recall that $z$ is in the interior of $\xx$, such that a ball of radius \smash{$\beta_h^{1/2}$}
centered at $z$ is contained inside $\xx$; thus, edge effects from $\xx_h$ being finite are negligible.

Meanwhile, by feasibility of $\tf^{(h)}_w(\cdot)$, we know that $|\Delta(r, \, m)| \leq \lambda_1\Norm{hv}_2^2 m(m-1)/2$.
From here, we can verify that
\begin{align*}
&\frac{h^k}{\beta_h^{k+2}} \sum_{r \in \rr}  \varphi_{k-1}\p{\frac{a_r}{\beta_h}} \sum_{m \in \set_r} \p{\frac{\p{\delta_r + mh}^2 \Norm{v}_2^2}{\beta_h^2} - 1}    \varphi_1\p{\frac{\p{\delta_r + mh} \Norm{v}_2}{\beta_h}} \Delta(r, \, m) \\
&\ \ \ \ \leq \lambda_1 \frac{h^k}{\beta_h^{k}} \sum_{r \in \rr} \varphi_{k-1}\p{\frac{a_r}{\beta_h}} \frac{1}{2}\sum_{m \in \set_r} \p{\frac{\Norm{hv}_2^4 m^4}{\beta_h^4} - \frac{\Norm{hv}_2^2 m^2}{\beta_h^2}} \varphi_1\p{\frac{\p{\delta_r + mh} v}{\beta_h}} \p{1 + \oo\p{h}} \\
&\ \ \ \ =  \lambda_1 \frac{h^k}{\beta_h^{k}} \sum_{r \in \rr} \varphi_{k-1}\p{\frac{a_r}{\beta_h}} \p{\frac{\beta_h}{h\Norm{v}_2} + o(1)} \\
&\ \ \ \ = \lambda_1\p{1 + o(1)}
\end{align*}
in the limit where $h \rightarrow 0$. To verify the last result, note that the density of points $a_r$ in the $k-1$ dimensional
normal space to $v$ is $\Norm{v}_2h^{1-k}$; we also recall that $\p{\EE{Z^4} - \EE{Z^2}}/2 = 1$ for a standard Gaussian random
variable $Z$.

\end{document}